\pdfoutput=1
\documentclass{amsart}
\usepackage{amsrefs}
\usepackage{hyperref}

\usepackage[labelformat=simple]{subcaption}

\makeatletter
\newcommand*{\da@rightarrow}{\mathchar
  "0\hexnumber@\symAMSa 4B\!\!}
\newcommand*{\da@leftarrow}{\mathchar"0\hexnumber@\symAMSa 4C }
\newcommand*{\xdashrightarrow}[2][]{%
  \mathrel{%
    \mathpalette{\da@xarrow{#1}{#2}{}\da@rightarrow{\,}{}}{}%
  }%
}
\newcommand{\xdashleftarrow}[2][]{%
  \mathrel{%
    \mathpalette{\da@xarrow{#1}{#2}\da@leftarrow{}{}{\,}}{}%
  }%
}
\newcommand*{\da@xarrow}[7]{%
  \sbox0{$\ifx#7\scriptstyle\scriptscriptstyle\else\scriptstyle\fi#5#1#6\m@th$}%
  \sbox2{$\ifx#7\scriptstyle\scriptscriptstyle\else\scriptstyle\fi#5#2#6\m@th$}%
  \sbox4{$#7\dabar@\m@th$}%
  \dimen@=\wd0 %
  \ifdim\wd2 >\dimen@
    \dimen@=\wd2 %
  \fi
  \count@=2 %
  \def\da@bars{\dabar@\dabar@}%
  \@whiledim\count@\wd4<\dimen@\do{%
    \advance\count@\@ne
    \expandafter\def\expandafter\da@bars\expandafter{%
      \da@bars
      \dabar@ 
    }%
  }%
  \mathrel{#3}%
  \mathrel{%
    \mathop{\da@bars}\limits
    \ifx\\#1\\%
    \else
      _{\copy0}%
    \fi
    \ifx\\#2\\%
    \else
      ^{\copy2}%
    \fi
  }%
  \mathrel{#4}%
}
\makeatother

\usepackage{tikz}
\usepackage{tikz-3dplot}
\usepackage{braids}

\usetikzlibrary{arrows,arrows.meta,braids}

\tikzset{
  baseline=(current bounding box.center),
  x=28pt, y=28pt, font=\small,
  >=stealth,
  vertex/.style={inner sep=0pt},
  lvertex/.style={vertex, draw, fill= white, opacity=1, minimum size=9pt, font=\footnotesize},
  gnode/.style={lvertex, shape=circle},
  fnode/.style={lvertex, shape=rectangle},
  q->/.style={->, shorten >=0.5pt},
  c->/.style={q->, thick, blue},
  f->/.style={q->, thick, red, densely dashed},
}

\newcommand{\CI}{\mathcal{I}}
\newcommand{\CN}{\mathcal{N}}
\newcommand{\cD}{\mathcal{D}}
\newcommand{\CA}{\mathcal{A}}
\newcommand{\CH}{\mathcal{H}}

\newcommand{\Z}{\mathbb{Z}}
\newcommand{\R}{\mathbb{R}}
\newcommand{\C}{\mathbb{C}}
\newcommand{\T}{\mathbb{T}}

\newcommand{\rmd}{\mathrm{d}}
\newcommand{\iu}{\mathrm{i}}

\newcommand{\U}{\mathrm{U}}
\newcommand{\SU}{\mathrm{SU}}
\newcommand{\SO}{\mathrm{SO}}

\newcommand{\ib}{{\bar{\imath}}}
\newcommand{\jb}{{\bar{\jmath}}}

\newcommand{\thetah}{{\hat{\theta}}}

\newcommand{\FS}{\mathfrak{S}}
\newcommand{\FSb}{\overline{\mathfrak{S}}}

\newcommand{\Lb}{\overline{L}}
\newcommand{\qb}{\bar{q}}
\newcommand{\Jb}{\overline{J}}

\newcommand{\sfX}{\mathsf{X}}
\newcommand{\sfY}{\mathsf{Y}}
\newcommand{\sfZ}{\mathsf{Z}}

\DeclareMathOperator*{\gauge}{Gauge}
\DeclareMathOperator*{\Res}{Res}
\DeclareMathOperator{\Tr}{Tr}
\let\Im\relax
\DeclareMathOperator{\Im}{Im}

\title{Tetrahedron duality}

\author{Junya Yagi}
\address{Yau Mathematical Sciences Center, Tsinghua University, China}

\date{}

\begin{document}
\begin{abstract}
  A certain two-dimensional supersymmetric gauge theory is argued to
  satisfy a relation that promotes the Zamolodchikov tetrahedron
  equation to an infrared duality between two quantum field theories.
  Solutions of the tetrahedron equation with continuous spin variables
  are obtained from partition functions of this theory and its
  variants.
\end{abstract}

\maketitle

\section{Introduction}
\label{sec:intro}

Dualities of supersymmetric gauge theories yield solutions of the
Yang--Baxter equation \cites{Bazhanov:2010kz, Bazhanov:2011mz,
  Spiridonov:2010em, Yamazaki:2012cp}.  In this paper we discuss
solutions of the Zamolodchikov tetrahedron equation \cite{MR611994b}
which similarly arise from gauge theory dualities.

\subsection{The tetrahedron equation  for IRC models}

The tetrahedron equation is a three-dimensional analog of the
Yang--Baxter equation and appears as a condition for integrability in
the context of classical spin models on three-dimensional lattices.

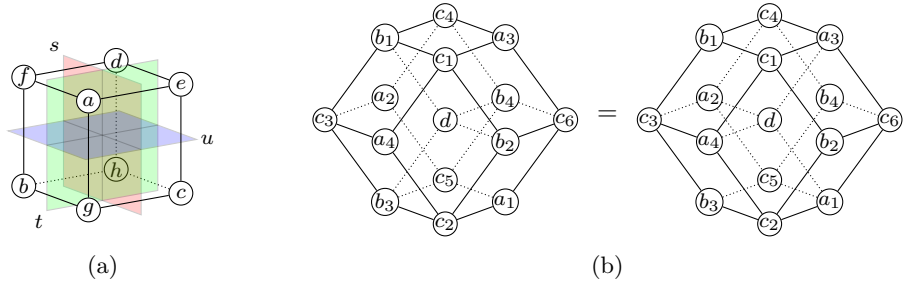
\begin{figure}
  \centering
  \begin{subfigure}{0.2\textwidth}
    \tdplotsetmaincoords{-75}{-35}
    \begin{tikzpicture}[tdplot_main_coords, scale=1.5, rotate=0, font=\footnotesize]
      \node[gnode] (b) at (0,1,1) {$b$};
      \node[gnode] (c) at (1,0,1) {$c$};
      \node[gnode] (d) at (1,1,0) {$d$};
      \node[gnode] (h) at (1,1,1) {$h$};
      
      \draw[densely dotted] (b) -- (h);
      \draw[densely dotted] (c) -- (h);
      \draw[densely dotted] (d) -- (h);
      
      \draw[fill=red, opacity=0.2]
      (0.5,-0.1,-0.1) -- (0.5,1.1,-0.1) node[above left=-2pt, opacity=1] {$s$}
      -- (0.5,1.1,1.1) -- (0.5,-0.1,1.1) -- cycle;
      
      \draw[fill=green, opacity=0.2]
      (-0.1,0.5,-0.1) -- (-0.1,0.5,1.1) node[below left=-2pt, opacity=1] {$t$}
      -- (1.1,0.5,1.1) -- (1.1,0.5,-0.1) -- cycle;
      
      \draw[fill=blue, opacity=0.2]
      (-0.1,-0.1,0.5) -- (1.1,-0.1,0.5) node[right=-2pt, opacity=1] {$u$}
      -- (1.1,1.1,0.5) -- (-0.1,1.1,0.5) -- cycle;
      
      \draw[opacity=0.2] (0.5,0.5,-0.1) -- (0.5,0.5,1.1);
      \draw[opacity=0.2] (0.5,-0.1,0.5) -- (0.5,1.1,0.5);
      \draw[opacity=0.2] (-0.1,0.5,0.5) -- (1.1,0.5,0.5);
      
      \node[gnode] (a) at (0,0,0) {$a$};
      \node[gnode] (e) at (1,0,0) {$e$};
      \node[gnode] (f) at (0,1,0) {$f$};
      \node[gnode] (g) at (0,0,1) {$g$};
      
      \draw (a) -- (e);
      \draw (a) -- (f);
      \draw (a) -- (g);
      \draw (e) -- (c);
      \draw (e) -- (d);
      \draw (f) -- (d);
      \draw (f) -- (b);
      \draw (g) -- (b);
      \draw (g) -- (c);
    \end{tikzpicture}
    
    \caption{}
    \label{fig:W}
  \end{subfigure}
  \qquad\qquad
  \begin{subfigure}{0.62\textwidth}
    \tdplotsetmaincoords{60}{270}
    \begin{tikzpicture}[tdplot_main_coords, scale=0.8]
    \node[gnode] (d) at (0,0,0) {$d$};

    \node[gnode] (a1) at (-1,-1,-1) {$a_1$};
    \node[gnode] (a2) at (-1,1,1) {$a_2$};
    \node[gnode] (a3) at (1,-1,1) {$a_3$};
    \node[gnode] (a4) at (1,1,-1) {$a_4$};
    
    \node[gnode] (b1) at (1,1,1) {$b_1$};
    \node[gnode] (b2) at (1,-1,-1) {$b_2$};
    \node[gnode] (b3) at (-1,1,-1) {$b_3$};
    \node[gnode] (b4) at (-1,-1,1) {$b_4$};

    \node[gnode] (c1) at (2,0,0) {$c_1$};
    \node[gnode] (c2) at (0,0,-2) {$c_2$};
    \node[gnode] (c3) at (0,2,0) {$c_3$};
    \node[gnode] (c4) at (0,0,2) {$c_4$};
    \node[gnode] (c5) at (-2,0,0) {$c_5$};
    \node[gnode] (c6) at (0,-2,0) {$c_6$};
    
    \draw[densely dotted] (b3) -- (c5);
    \draw (b3) -- (c3);
    \draw (b3) -- (c2);
    \draw (b1) -- (c4);
    \draw (b1) -- (c1);
    \draw (b1) -- (c3);
    \draw (b2) -- (c6);
    \draw (b2) -- (c1);
    \draw (b2) -- (c2);
    \draw[densely dotted] (b4) -- (c6);
    \draw[densely dotted] (b4) -- (c5);
    \draw[densely dotted] (b4) -- (c4);
    \draw (a3) -- (c6);
    \draw (a3) -- (c4);
    \draw (a3) -- (c1);
    \draw (a1) -- (c6);
    \draw[densely dotted] (a1) -- (c5);
    \draw (a1) -- (c2);
    \draw[densely dotted] (a2) -- (c5);
    \draw[densely dotted] (a2) -- (c4);
    \draw[densely dotted] (a2) -- (c3);
    \draw (a4) -- (c1);
    \draw (a4) -- (c3);
    \draw (a4) -- (c2);
    \draw[densely dotted] (b3) -- (d);
    \draw[densely dotted] (b1) -- (d);
    \draw[densely dotted] (b2) -- (d);
    \draw[densely dotted] (b4) -- (d);
  \end{tikzpicture}
  $\ \ = \ \ $
    \begin{tikzpicture}[tdplot_main_coords, scale=0.8]
    \node[gnode] (d) at (0,0,0) {$d$};

    \node[gnode] (a1) at (-1,-1,-1) {$a_1$};
    \node[gnode] (a2) at (-1,1,1) {$a_2$};
    \node[gnode] (a3) at (1,-1,1) {$a_3$};
    \node[gnode] (a4) at (1,1,-1) {$a_4$};
    
    \node[gnode] (b1) at (1,1,1) {$b_1$};
    \node[gnode] (b2) at (1,-1,-1) {$b_2$};
    \node[gnode] (b3) at (-1,1,-1) {$b_3$};
    \node[gnode] (b4) at (-1,-1,1) {$b_4$};

    \node[gnode] (c1) at (2,0,0) {$c_1$};
    \node[gnode] (c2) at (0,0,-2) {$c_2$};
    \node[gnode] (c3) at (0,2,0) {$c_3$};
    \node[gnode] (c4) at (0,0,2) {$c_4$};
    \node[gnode] (c5) at (-2,0,0) {$c_5$};
    \node[gnode] (c6) at (0,-2,0) {$c_6$};
    
    \draw[densely dotted] (b3) -- (c5);
    \draw (b3) -- (c3);
    \draw (b3) -- (c2);
    \draw (b1) -- (c4);
    \draw (b1) -- (c1);
    \draw (b1) -- (c3);
    \draw (b2) -- (c6);
    \draw (b2) -- (c1);
    \draw (b2) -- (c2);
    \draw[densely dotted] (b4) -- (c6);
    \draw[densely dotted] (b4) -- (c5);
    \draw[densely dotted] (b4) -- (c4);
    \draw (a3) -- (c6);
    \draw (a3) -- (c4);
    \draw (a3) -- (c1);
    \draw (a1) -- (c6);
    \draw[densely dotted] (a1) -- (c5);
    \draw (a1) -- (c2);
    \draw[densely dotted] (a2) -- (c5);
    \draw[densely dotted] (a2) -- (c4);
    \draw[densely dotted] (a2) -- (c3);
    \draw (a4) -- (c1);
    \draw (a4) -- (c3);
    \draw (a4) -- (c2);
    \draw[densely dotted] (a3) -- (d);
    \draw[densely dotted] (a1) -- (d);
    \draw[densely dotted] (a2) -- (d);
    \draw[densely dotted] (a4) -- (d);
  \end{tikzpicture}
    
    \caption{}
    \label{fig:TE-IRC}
  \end{subfigure}
  \caption{(a) The configuration of spin variables for the Boltzmann
    weight~\eqref{eq:W}.  (b) The tetrahedron equation in the IRC
    form.}
  \label{fig:IRC}
\end{figure}

In an interaction-round-a-cube (IRC) model, spin variables are located
at the vertices of a simple cubic lattice and their interactions take
place inside the cubes of the lattice.  The Boltzmann weight for the
interaction in a single cube is a function
\begin{equation}
  \label{eq:W}
  W(a|e,f,g|b,c,d|h; s, t, u)
\end{equation}
of eight spin variables $a$, $b$, \dots, $h$ at the vertices of the
cube.  It may also depend on parameters $s$, $t$, $u$, called the
\emph{spectral parameters}, which we think of as assigned to three
planes bisecting the cube.  See Figure~\ref{fig:W}.

The tetrahedron equation for IRC models reads \cite{MR696804}
\begin{multline}
  \label{eq:TE}
  \begin{aligned}
    \int_d &\mu(d)
    W(a_4|c_2,c_1,c_3|b_1,b_3,b_2|d; s_1, s_3, s_2)
    W(c_1|b_2,a_3,b_1|c_4,d,c_6|b_4; s_1, s_4, s_2)
    \\
    &
    \times
    W(b_1|d,c_4,c_3|a_2,b_3,b_4|c_5; s_1, s_4, s_3)
    W(d|b_2,b_4,b_3|c_5,c_2,c_6|a_1; s_2, s_4, s_3)
  \end{aligned}
  \\
  =
  \begin{aligned}
    \int_d & \mu(d)
    W(b_1|c_1,c_4,c_3|a_2,a_4,a_3|d; s_2, s_4, s_3)
    W(c_1|b_2,a_3,a_4|d,c_2,c_6|a_1; s_1, s_4, s_3)
    \\
    & \times
    W(a_4|c_2,d,c_3|a_2,b_3,a_1|c_5; s_1, s_4, s_2)
    W(d|a_1,a_3,a_2|c_4,c_5,c_6|b_4; s_1, s_3, s_2)
    \,,
  \end{aligned}
\end{multline}
where the integrals are performed for the value of the spin variable
$d$ with respect to a measure $\mu(d)$. Geometrically, the two sides
of the tetrahedron equation represent two ways of decomposing a
rhombic dodecahedron into four cubes, as depicted in
Figure~\ref{fig:TE-IRC}.

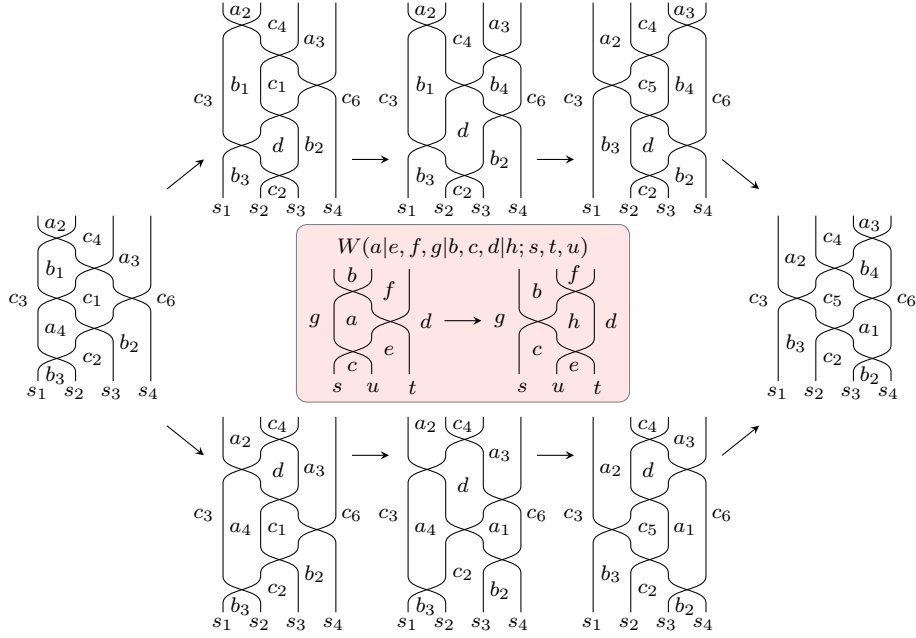
\begin{figure}
  \centering
  \begin{tikzpicture}[xscale=0.5, yscale=0.4, font=\footnotesize]
    \begin{scope}[shift={(8,-2)}]
      \draw[rounded corners, draw=black!50, fill=red!10] (0,1.5) rectangle (9,-4.5);

      \node at (4.5,0.7) {$W(a|e,f,g|b,c,d|h; s, t, u)$};

      \begin{scope}
        \braid[gap=0] a_1 a_2 a_1;
        
        \node at (1.5,-4) [left] {$s$};
        \node at (2.5,-4) [left] {$u$};
        \node at (3.5,-4) [left] {$t$};
        
        \node at (0.5,-1.75) {$g$};
        \node at (1.5,-1.75) {$a$};
        \node at (1.5,-0.2) {$b$};
        \node at (1.5,-3.3) {$c$};
        \node at (2.5,-2.75) {$e$};
        \node at (2.5,-0.75) {$f$};
        \node at (3.5,-1.75) {$d$};
      \end{scope}
      \draw[->, shift={(4,-1.75)}] (0,0) -- (1,0);
      \begin{scope}[shift={(5,0)}]
        \braid[gap=0] a_2 a_1 a_2;
        
        \node at (1.5,-4) [left] {$s$};
        \node at (2.5,-4) [left] {$u$};
        \node at (3.5,-4) [left] {$t$};
        
        \node at (0.5,-1.75) {$g$};
        \node at (2.5,-1.75) {$h$};
        \node at (1.5,-0.75) {$b$};
        \node at (1.5,-2.75) {$c$};
        \node at (2.5,-3.3) {$e$};
        \node at (2.5,-0.2) {$f$};
        \node at (3.5,-1.75) {$d$};
      \end{scope}
    \end{scope}

    \begin{scope}[shift={(0,-0.2)}]
      \braid[gap=0] a_1 a_2 a_3-a_1 a_2 a_1;
        
        \node at (1,-6) {$s_1$};
        \node at (2,-6) {$s_2$};
        \node at (3,-6) {$s_3$};
        \node at (4,-6) {$s_4$};
        
        \node at (0.5,-2.85) {$c_3$};
        
        \node at (1.5,-0.3) {$a_2$};
        \node at (1.5,-3.85) {$a_4$};
        \node at (1.5,-1.8) {$b_1$};
        \node at (1.5,-5.4) {$b_3$};
        
        \node at (2.5,-0.8) {$c_4$};
        \node at (2.5,-2.85) {$c_1$};
        \node at (2.5,-4.85) {$c_2$};
        
        \node at (3.5,-1.5) {$a_3$};
        \node at (3.5,-4.3) {$b_2$};
        
        \node at (4.5,-2.85) {$c_6$};
      \end{scope}

      \begin{scope}[shift={(5,7)}] 
        \braid[gap=0] a_1 a_2 a_3 a_2 a_1 a_2;
        
        \node at (1,-7) {$s_1$};
        \node at (2,-7) {$s_2$};
        \node at (3,-7) {$s_3$};
        \node at (4,-7) {$s_4$};
        
        \node at (0.5,-3.3) {$c_3$};
        
        \node at (1.5,-0.3) {$a_2$};
        \node at (1.5,-2.8) {$b_1$};
        \node at (1.5,-5.85) {$b_3$};
        
        \node at (2.5,-0.8) {$c_4$};
        \node at (2.5,-2.8) {$c_1$};
        \node at (2.5,-6.4) {$c_2$};
        \node at (2.5,-4.8) {$d$};
        
        \node at (3.5,-1.3) {$a_3$};
        \node at (3.5,-4.8) {$b_2$};
        
        \node at (4.5,-3.3) {$c_6$};
      \end{scope}

      \begin{scope}[shift={(10,7)}] 
        \braid[gap=0] a_1 a_3 a_2 a_3 a_1 a_2;
        
        \node at (1,-7) {$s_1$};
        \node at (2,-7) {$s_2$};
        \node at (3,-7) {$s_3$};
        \node at (4,-7) {$s_4$};
        
        \node at (0.5,-3.3) {$c_3$};
        
        \node at (1.5,-0.3) {$a_2$};
        \node at (1.5,-2.8) {$b_1$};
        \node at (1.5,-5.85) {$b_3$};
        
        \node at (2.5,-1.3) {$c_4$};
        \node at (2.5,-6.4) {$c_2$};
        \node at (2.5,-4.3) {$d$};
        
        \node at (3.5,-0.8) {$a_3$};
        \node at (3.5,-2.8) {$b_4$};
        \node at (3.5,-5.3) {$b_2$};
        
        \node at (4.5,-3.3) {$c_6$};
      \end{scope}

      \begin{scope}[shift={(15,7)}] 
        \braid[gap=0] a_3 a_2 a_1 a_2 a_3 a_2;
        
        \node at (1,-7) {$s_1$};
        \node at (2,-7) {$s_2$};
        \node at (3,-7) {$s_3$};
        \node at (4,-7) {$s_4$};
        
        \node at (0.5,-3.3) {$c_3$};
        
        \node at (1.5,-1.3) {$a_2$};
        \node at (1.5,-4.8) {$b_3$};
        
        \node at (2.5,-0.8) {$c_4$};
        \node at (2.5,-2.8) {$c_5$};
        \node at (2.5,-4.8) {$d$};
        \node at (2.5,-6.4) {$c_2$};
        
        \node at (3.5,-0.3) {$a_3$};
        \node at (3.5,-2.8) {$b_4$};
        \node at (3.5,-5.85) {$b_2$};
        
        \node at (4.5,-3.3) {$c_6$};
      \end{scope}

      \begin{scope}[shift={(5,-7)}] 
        \braid[gap=0] a_2 a_1 a_2 a_3 a_2 a_1;
        
        \node at (1,-7) {$s_1$};
        \node at (2,-7) {$s_2$};
        \node at (3,-7) {$s_3$};
        \node at (4,-7) {$s_4$};
        
        \node at (0.5,-3.3) {$c_3$};
        
        \node at (1.5,-0.8) {$a_2$};
        \node at (1.5,-3.8) {$a_4$};
        \node at (1.5,-6.4) {$b_3$};
        
        \node at (2.5,-0.3) {$c_4$};
        \node at (2.5,-1.8) {$d$};
        \node at (2.5,-3.8) {$c_1$};
        \node at (2.5,-5.85) {$c_2$};
        
        \node at (3.5,-1.8) {$a_3$};
        \node at (3.5,-5.3) {$b_2$};
        
        \node at (4.5,-3.3) {$c_6$};
      \end{scope}

      \begin{scope}[shift={(10,-7)}] 
        \braid[gap=0] a_2 a_1 a_3 a_2 a_3 a_1;
        
        \node at (1,-7) {$s_1$};
        \node at (2,-7) {$s_2$};
        \node at (3,-7) {$s_3$};
        \node at (4,-7) {$s_4$};
        
        \node at (0.5,-3.3) {$c_3$};
        
        \node at (1.5,-0.3) {$a_2$};
        \node at (1.5,-3.8) {$a_4$};
        \node at (1.5,-6.4) {$b_3$};
        
        \node at (2.5,-0.3) {$c_4$};
        \node at (2.5,-2.3) {$d$};
        \node at (2.5,-5.3) {$c_2$};
        
        \node at (3.5,-1.3) {$a_3$};
        \node at (3.5,-3.8) {$a_1$};
        \node at (3.5,-5.85) {$b_2$};
        
        \node at (4.5,-3.3) {$c_6$};
      \end{scope}

      \begin{scope}[shift={(15,-7)}] 
        \braid[gap=0] a_2 a_3 a_2 a_1 a_2 a_3;
        
        \node at (1,-7) {$s_1$};
        \node at (2,-7) {$s_2$};
        \node at (3,-7) {$s_3$};
        \node at (4,-7) {$s_4$};
        
        \node at (0.5,-3.3) {$c_3$};
        
        \node at (1.5,-1.8) {$a_2$};
        \node at (1.5,-5.3) {$b_3$};
        
        \node at (2.5,-0.3) {$c_4$};
        \node at (2.5,-1.8) {$d$};
        \node at (2.5,-3.8) {$c_5$};
        \node at (2.5,-5.85) {$c_2$};
        
        \node at (3.5,-0.8) {$a_3$};
        \node at (3.5,-3.8) {$a_1$};
        \node at (3.5,-6.4) {$b_2$};
        
        \node at (4.5,-3.3) {$c_6$};
      \end{scope}

      \begin{scope}[shift={(20,-0.2)}] 
        \braid[gap=0] a_3 a_2 a_3-a_1 a_2 a_3;
        
        \node at (1,-6) {$s_1$};
        \node at (2,-6) {$s_2$};
        \node at (3,-6) {$s_3$};
        \node at (4,-6) {$s_4$};
        
        \node at (0.5,-2.85) {$c_3$};
        
        \node at (1.5,-1.5) {$a_2$};
        \node at (1.5,-4.3) {$b_3$};
        
        \node at (2.5,-0.8) {$c_4$};
        \node at (2.5,-2.85) {$c_5$};
        \node at (2.5,-4.85) {$c_2$};
        
        \node at (3.5,-0.3) {$a_3$};
        \node at (3.5,-1.8) {$b_4$};
        \node at (3.5,-3.85) {$a_1$};
        \node at (3.5,-5.4) {$b_2$};
        
        \node at (4.5,-2.85) {$c_6$};
      \end{scope}

      \draw[->, shift={(4.5,4)}] (0,-3.3) -- (1,-2.3);
      \draw[->, shift={(4.5,-4)}] (0,-3.3) -- (1,-4.3);

      \draw[->, shift={(9.5,5)}] (0,-3.3) -- (1,-3.3);
      \draw[->, shift={(9.5,-5)}] (0,-3.3) -- (1,-3.3);

      \draw[->, shift={(14.5,5)}] (0,-3.3) -- (1,-3.3);
      \draw[->, shift={(14.5,-5)}] (0,-3.3) -- (1,-3.3);

      \draw[->, shift={(19.5,4)}] (0,-2.3) -- (1,-3.3);
      \draw[->, shift={(19.5,-4)}] (0,-4.3) -- (1,-3.3);
  \end{tikzpicture}
  \caption{The presentations of the Boltzmann weight \eqref{eq:W} and
    the tetrahedron equation \eqref{eq:TE} using curves in a plane.
    Spin variables are assigned to the regions separated by the
    curves.}
  \label{fig:TE-wires}
\end{figure}

The Boltzmann weight \eqref{eq:W} may also be presented as a
\emph{Yang--Baxter move} (or \emph{braid move}) on a configuration of
three curves in a plane.  In Figure~\ref{fig:W}, imagine a fourth
plane that is perpendicular to the line through $a$ and $h$.  Its
intersections with the other three planes make three straight lines.
As the fourth plane moves from the front to the back side of the cube,
the three lines undergo a Yang--Baxter move.  The tetrahedron equation
amounts to an equality between two sequences of four Yang--Baxter
moves applied to a configuration of four curves.  See
Figure~\ref{fig:TE-wires}.

\subsection{Solution by the brane box model}

The solutions of the tetrahedron equation \eqref{eq:TE} discussed in
this paper originate from a two-dimensional gauge theory with
$\CN = (0,2)$ supersymmetry as well as its variants.  This gauge
theory, known as the \emph{brane box model} \cites{Mohri:1997ef,
  GarciaCompean:1998kh}, is realized by intersecting branes that form
a cubic lattice in a three-dimensional subspace of the ten-dimensional
spacetime of string theory.

\begin{figure}
  \tdplotsetmaincoords{-70}{-37}
  \begin{subfigure}{0.3\textwidth}
  \begin{tikzpicture}[tdplot_main_coords, scale=1.2, rotate=0]
    \foreach \l [count=\li] in {0,...,2}
    \foreach \m [count=\mi] in {0,...,2}
    \foreach \n [count=\ni] in {0,...,2}
    \node[gnode, minimum size=3pt] (\l\m\n) at (\l,\m,\n) {};

    \foreach \l [count=\li] in {0,...,1}
    \foreach \m in {0,...,2}
    \foreach \n in {0,...,2}
    \draw[c->, thin] (\l\m\n) -- (\li\m\n);

    \foreach \l in {0,...,2}
    \foreach \m [count=\mi] in {0,...,1}
    \foreach \n in {0,...,2}
    \draw[c->, thin] (\l\m\n) -- (\l\mi\n);

    \foreach \l in {0,...,2}
    \foreach \m in {0,...,2}
    \foreach \n [count=\ni] in {0,...,1}
    \draw[c->, thin] (\l\m\n) -- (\l\m\ni);

    \foreach \l [count=\li] in {0,...,1}
    \foreach \m [count=\mi] in {0,...,1}
    \foreach \n [count=\ni] in {0,...,1}
    \draw[c->, thin] (\li\mi\ni) -- (\l\m\n);

    \foreach \l in {0,...,2}
    \foreach \m [count=\mi] in {0,...,1}
    \foreach \n [count=\ni] in {0,...,1}
    \draw[f->, thin] (\l\m\n) -- (\l\mi\ni);

    \foreach \l [count=\li] in {0,...,1}
    \foreach \m in {0,...,2}
    \foreach \n [count=\ni] in {0,...,1}
    \draw[f->, thin] (\l\m\n) -- (\li\m\ni);

    \foreach \l [count=\li] in {0,...,1}
    \foreach \m [count=\mi] in {0,...,1}
    \foreach \n in {0,...,2}
    \draw[f->, thin] (\l\m\n) -- (\li\mi\n);
  \end{tikzpicture}

  \caption{}
  \label{fig:BBM}
  \end{subfigure}
  \qquad
    \tikzset{gnode/.style={lvertex, shape=circle, minimum size=6pt}}
  \begin{subfigure}{0.2\textwidth}
    \begin{tikzpicture}[tdplot_main_coords, scale=1.5]
      \node[gnode] (a) at (0,0,0) {};
      \node[gnode] (e) at (1,0,0) {};
      \node[gnode] (f) at (0,1,0) {};
      \node[gnode] (g) at (0,0,1) {};
      \node[gnode] (b) at (0,1,1) {};
      \node[gnode] (c) at (1,0,1) {};
      \node[gnode] (d) at (1,1,0) {};
      \node[gnode] (h) at (1,1,1) {};
      
      \draw[densely dotted] (b) -- (h);
      \draw[densely dotted] (c) -- (h);
      \draw[densely dotted] (d) -- (h);
      
      \draw (a) -- (e);
      \draw (a) -- (f);
      \draw (a) -- (g);
      \draw (e) -- (c);
      \draw (e) -- (d);
      \draw (f) -- (d);
      \draw (f) -- (b);
      \draw (g) -- (b);
      \draw (g) -- (c);

      \draw[f->] (a) -- (b);
      \draw[f->] (a) -- (c);
      \draw[f->] (a) -- (d);
      \draw[c->] (a) -- (e);
      \draw[c->] (a) -- (f);
      \draw[c->] (a) -- (g);
      \draw[c->] (h) -- (a);
    \end{tikzpicture}
    
    \caption{}
    \label{fig:NUC}
  \end{subfigure}
  \qquad
  \begin{subfigure}{0.2\textwidth}
  \begin{tikzpicture}[tdplot_main_coords, scale=1.5]
    \node[gnode] (a) at (0,0,0) {};
    \node[gnode] (e) at (1,0,0) {};
    \node[gnode] (f) at (0,-1,0) {};
    \node[gnode] (b) at (0,0,1) {};
    \node[gnode] (g) at (0,1,1) {};
    \node[gnode] (c) at (1,1,1) {};
    \node[gnode] (d) at (1,-1,0) {};
    \node[gnode] (h) at (1,0,1) {};

    \draw[f->] (a) -- (h);
    \draw[f->] (b) -- (c);
    \draw[f->] (e) -- (c);
    \draw[c->] (a) -- (e);
    \draw[c->] (a) -- (b);
    \draw[c->] (c) -- (a);
    \draw[c->] (h) -- (c);

    \draw (a) -- (f);
    \draw (a) -- (g);
    \draw (e) -- (d);
    \draw (f) -- (d);
    \draw (f) -- (b);
    \draw (g) -- (b);
    \draw[densely dotted] (g) -- (c);
    \draw (b) -- (h);
    \draw (d) -- (h);

    \node[gnode, draw=black!20] (1) at (0,1,0) {};
    \node[gnode, draw=black!20] (2) at (1,1,0) {};
    \node[gnode, draw=black!20] (3) at (0,-1,1) {};
    \node[gnode, draw=black!20] (4) at (1,-1,1) {};

    \draw[-, draw=black!20] (a) -- (1) -- (2) -- (e) (1) -- (g) (f) -- (3) -- (4) -- (d) (b) -- (3);
    \draw[-, draw=black!20, densely dotted] (2) -- (c) (e) -- (h) (4) -- (h);
  \end{tikzpicture}
    
    \caption{}
    \label{fig:OUC}
  \end{subfigure}

  \caption{(a) Part of the brane box model that contains
    $2 \times 2 \times 2$ cubes.  (b) The natural unit cell of the
    brane box model.  (c) The oblique unit cell which satisfies the
    tetrahedron duality.}
\end{figure}
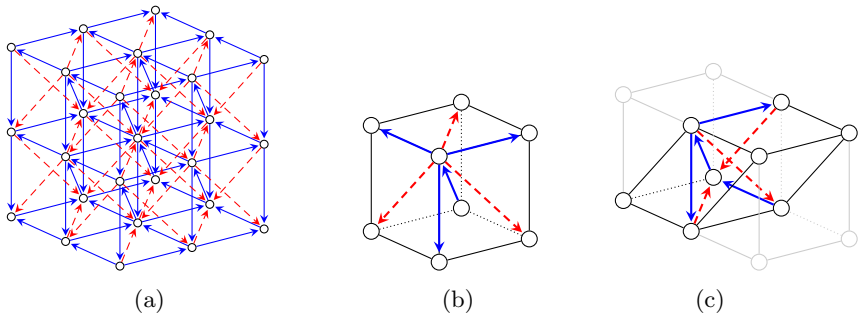

The gauge group and the matter content of the brane box model may be
presented as if they constitute a spin model analogous to the Ising
model.  The ``spins'' are located at the vertices of a simple cubic
lattice, and each corresponds to a $\U(N)$ factor in the gauge group.
There are two types of pairwise ``interactions'' (or ``bonds'')
between spins, produced by chiral multiplets and Fermi multiplets in
bifundamental representations.%
\footnote{The actual interactions in the brane
box model as a field theory are due to coupling to a vector multiplet
and a cubic superpotential.}
Part of this spin model is illustrated in Figure~\ref{fig:BBM}, where
spins are drawn as circles and interactions are represented by solid
and dashed arrows.

Given a spin model on a lattice, the choice of unit cell from which
the model is constructed is far from unique.  In the case of the brane
box model, however, there is a unique unit cell of smallest size whose
faces are parallel to branes in the underlying string theory picture.%
\footnote{There is still a choice to be made regarding the locations
  of seven arrows in the unit cell.}
This unit cell is shown in Figure~\ref{fig:NUC}.  It is a cube of unit
volume with eight spins located at its vertices and represents a
theory with eight $\U(N)$ flavor symmetry%
\footnote{A \emph{flavor symmetry} of a supersymmetric field theory is
  a global symmetry that commutes with the supersymmetry of the
  theory.}
groups.  The brane box model is obtained from infinite copies of this
theory glued together.  The gluing is done by gauging: to glue two
copies of the unit cell, one identifies the $\U(N)$ symmetries at the
overlapping vertices of the two cubes and changes them to gauge
symmetries.

In \cite{Yagi:2015lha}, the author argued that the above natural
choice of unit cell should lead to a solution of the tetrahedron
equation.  The argument adapted a beautiful idea of Kevin Costello
\cites{Costello:2013zra, Costello:2013sla} and involves surface defects
in a partially topological quantum field theory.  Although the
reasoning was seemingly solid, there was one problem: the natural unit
cell did not appear to satisfy the tetrahedron equation!

In the present work we partially resolve this problem.  Consider
another unit cell, which is \emph{slanted} compared to the natural
unit cell by one unit along one of the three axes of the cubic
lattice, as shown in Figure~\ref{fig:OUC}.  We will call it the
\emph{oblique unit cell} of the brane box model.  Under a reasonable
assumption about the cancellation of gauge anomaly, we will show that
the oblique unit cell does satisfy the tetrahedron equation.

\subsection{The tetrahedron duality}

The statement ``the oblique unit cell satisfies the tetrahedron
equation'' needs to be made more precise.

Let
\begin{equation}
  \label{eq:OUC}
  (a|e,f,g|b,c,d|h; s, t, u)
\end{equation}
denote the theory represented by the oblique unit cell.  This is an
$\CN = (0,2)$ supersymmetric field theory with eight $\U(N)$ flavor
symmetry groups, $\U(N)_a$, $\U(N)_b$, \dots, $\U(N)_h$, and the
R-charge assignment for its matter content depends on spectral
parameters $s$, $t$, $u$.

The claim is that the oblique unit cell \eqref{eq:OUC} satisfies the
\emph{tetrahedron duality}
\begin{multline}
  \label{eq:TD}
  \gauge_d
  \left(
    \begin{aligned}
      &
      (a_4|c_2,c_1,c_3|b_1,b_3,b_2|d; s_1, s_3, s_2)
      (c_1|b_2,a_3,b_1|c_4,d,c_6|b_4; s_1, s_4, s_2)
      \\
      & 
      (b_1|d,c_4,c_3|a_2,b_3,b_4|c_5; s_1, s_4, s_3)
      (d|b_2,b_4,b_3|c_5,c_2,c_6|a_1; s_2, s_4, s_3)
    \end{aligned}
  \right)
  \\
  \simeq
  \gauge_d
  \left(
    \begin{aligned}
      &
      (b_1|c_1,c_4,c_3|a_2,a_4,a_3|d; s_2, s_4, s_3)
      (c_1|b_2,a_3,a_4|d,c_2,c_6|a_1; s_1, s_4, s_3)
      \\
      &
      (a_4|c_2,d,c_3|a_2,b_3,a_1|c_5; s_1, s_4, s_2)
      (d|a_1,a_3,a_2|c_4,c_5,c_6|b_4; s_1, s_3, s_2)
    \end{aligned}
  \right)
  \,,
\end{multline}
where $\gauge_d$ is the operation of gauging $\U(N)_d$ and $\simeq$
means equivalence of theories at low energies.  Each side of this
relation is an $\CN = (0,2)$ supersymmetric gauge theory constructed
from four copies of the oblique unit cell coupled by gauging
$\U(N)_d$.  The two sides have the same gauge group and global
symmetry groups, but their matter contents are different.

In fact, the tetrahedron duality \eqref{eq:TD} is a consequence of the
\emph{$(0,2)$ triality}, an infrared equivalence of a triplet of
$\CN = (0,2)$ supersymmetric gauge theories with unitary gauge groups
\cite{Gadde:2013lxa}, again under a reasonable assumption about
anomaly cancellation.

\subsection{Elliptic genus solutions}

Since the tetrahedron duality \eqref{eq:TD} holds at the level of
quantum field theory, any physical quantity computed at low energies
in the theories represented by the two sides match.  If a quantity $W$
is compatible with the gauging operation in the sense that it
satisfies the relation
\begin{equation}
  W \gauge_d \simeq \int_d \mu(d) W \,,
\end{equation}
then $W$ computed in the oblique unit cell \eqref{eq:OUC} (for which
we will use the notation~\eqref{eq:W}) is a solution of the
tetrahedron equation \eqref{eq:TE}.  The partition function of the
lattice model defined by this solution equals $W$ computed in the
brane box model.

Elliptic genera are excellent examples of such quantities.  There are
different versions of elliptic genera, but all of them are partition
functions on a torus $\C/(\Z + \tau \Z)$, $\Im \tau > 0$, and take the
form
\begin{equation}
  \Tr_{\CH_{\text{BPS}}}\biggl(
  (-1)^F q^{H_L} \prod_\alpha u_\alpha^{K_\alpha}
  \biggr) \,.
\end{equation}
Here $\CH_{\text{BPS}}$ is a space of states preserving half of
$\CN = (0,2)$ supersymmetry, $(-1)^F$ is a fermion parity operator,
$H_L$ is a left-moving Hamiltonian and $q = e^{2\pi\iu\tau}$; the
theory has $\U(1)$ flavor symmetries $\U(1)_\alpha$, $\alpha = 1$,
\dots, $N_f$, $K_\alpha$ is a generator of $\U(1)_\alpha$ and
$u_\alpha$ is an associated parameter, called the \emph{fugacity} for
$K_\alpha$.  Being a partition function on an elliptic curve that can
be written as a trace, an elliptic genus is a power series in $q$ and
$u_\alpha$ with integer coefficients and has a nice modular property.

Not only are elliptic genera compatible with gauging, but they are
also invariant under continuous deformations of the theory (barring
wall-crossing phenomena).  As such, they can be computed exactly in
the weak coupling limit, and integral formulas are known for theories
with standard Lagrangian descriptions \cites{Benini:2013nda,
  Benini:2013xpa}.

\subsection{Abelian gauge anomalies}

The elliptic genera of the brane box model thus provide solutions of
the tetrahedron equation \eqref{eq:TE}.  There is a caveat, however.

For the brane box model to define a consistent quantum field theory,
the gauge anomaly in the abelian part of the gauge group, generated by
chiral fermions, must be canceled.  The theory represented by the
oblique unit cell is also plagued by an abelian gauge anomaly.  A
version of the Green--Schwarz mechanism is expected to cancel these
anomalies \cite{Mohri:1997ef}.

The $(0,2)$ triality, on the other hand, has been mostly supported by
calculations using another method of cancelling abelian gauge
anomalies.  In this method, additional Fermi multiplets in determinant
representations are introduced.

In the claim made above that the oblique unit cell satisfies the
tetrahedron duality, we have assumed that \emph{the cancellation of
  abelian gauge anomaly in the brane box model does not compromise the
  validity of the $(0,2)$ triality}.  There is a good reason to
believe that this is the case based on the study of brane brick models
\cites{Franco:2015tna, Franco:2015tya, Franco:2016nwv,
  Franco:2017cjj}, of which the brane box model is one example.

That said, the precise mechanism of anomaly cancellation for the brane
box model is yet to be understood.  This fact prevents us from writing
down the elliptic genus solutions.

\subsection{Solutions by the 1D brane box model and the
  \texorpdfstring{$(0,1)$}{(0,2)} brane box model}

The problem stemming from the lack of understanding of anomaly
cancellation can be circumvented in two ways:%
\footnote{Here are some of the failed attempts that the author has
  made to cancel the abelian gauge anomaly and write down the elliptic
  genus solutions: (1) Replace $\U(N)$ gauge groups with $\SU(N)$.
  (There is no triality for $\SU(N)$ gauge groups.)  (2) Introduce
  extra Fermi multiplets.  (It seems hard to cancel the anomaly while
  preserving the locality of the lattice model, let alone the $(0,2)$
  triality.)  (3) Add a determinant Fermi multiplet for every
  bifundamental chiral multiplet and a determinant chiral multiplet
  for every bifundamental Fermi multiplet.  (The additional multiplets
  are not compatible with the $(0,2)$ triality.)}
\begin{enumerate}
\item Dimensionally reduce the brane box model to one dimension.  We
  will call the resulting theory the \emph{1D brane box model}.  Since
  chiral anomalies are absent in odd dimensions, the abelian gauge
  anomaly vanishes after the dimensional reduction.  The oblique unit
  cell for the 1D brane box model satisfies the tetrahedron duality
  and its Witten genus (which is a partition function on a circle) is
  a solution of the tetrahedron equation.

\item Replace the brane box model with an $\CN = (0,1)$ supersymmetric
  gauge theory with special orthogonal gauge group.  This theory,
  which we will refer to as the \emph{$(0,1)$ brane box model}, is
  free of gauge anomaly and enjoys a triality almost identical to the
  $(0,2)$ triality \cite{Gukov:2019lzi}.  Consequently, the oblique
  unit cell for the $(0,1)$ brane box model satisfies the tetrahedron
  duality and its elliptic genera solve the tetrahedron equation,
  though they lack spectral parameters since there is no continuous
  R-symmetry for $\CN = (0,1)$ supersymmetry.
\end{enumerate}

We will write down two solutions of the tetrahedron equations produced
by the 1D brane box model and the $(0,1)$ brane box model.  The
lattice models defined by these solutions have continuous spin
variables, coming from the fugacities for the flavor symmetry groups
of the oblique unit cells.

\subsection{Open questions}

While we obtain solutions of the tetrahedron equation from the brane
box model and related theories, we leave many questions unanswered in
this paper.  Some important ones are:
\begin{itemize}
\item Why does the unit cell need to be oblique in order for it
  to satisfy the tetrahedron duality?

\item How is the abelian gauge anomaly canceled in the brane box
  model?

\item What is the significance of the third theory related by the
  triality to the two sides of the tetrahedron duality?
\end{itemize}

Seeking connections with other works, we can further ask interesting
questions.  Here are two:
\begin{itemize}
\item String theory realizations of the trialities of brane brick
  models and their $\CN = (0,1)$ supersymmetric projections
  \cite{Franco:2021ixh} were given in \cite{Franco:2016nwv} and
  \cite{Franco:2021vxq}.  Do brane brick models provide more examples
  of theories satisfying the tetrahedron duality?

\item Solutions of the tetrahedron equation were constructed in
  \cites{Sun:2022mpy, Inoue:2023vtx, Inoue:2023rer, Inoue:2024swb} by
  a method using quantum cluster algebras.  They are closely related
  to the solution \cites{MR1278735, Bazhanov:2005as} realized by a
  brane configuration in M-theory \cite{Yagi:2022tot}.  What are the
  relations, if any, of these solutions to the ones obtained in this
  paper?
\end{itemize}
In connection with the former question, it should be mentioned that in
\cite{Franco:2016nwv} a rhombic dodecahedron structure was observed in
brane brick models and a possibility of its relevance to the
tetrahedron equation was suggested.

\section{The brane box model and the tetrahedron duality}
\label{sec:BBM-TD}

In this section we introduce the brane box model and explain how the
$(0,2)$ triality implies that its oblique unit cell satisfies the
tetrahedron duality.  We also discuss the cancellation of abelian
gauge anomaly and the relation of spectral parameters to boundary
conditions on the lattice.

\subsection{The brane box model}
\label{sec:BBM}

Consider a three-dimensional Euclidean subspace $\R^3$ of the
ten-dimensional Minkowski spacetime of type IIA string theory.  Let
$(x,y,z)$ be the standard coordinates of $\R^3$ and
\begin{equation}
  \label{eq:XYZ}
  \sfX_l = \{x = l + \tfrac12\} \,,
  \quad
  \sfY_m = \{y = m + \tfrac12\} \,,
  \quad
  \sfZ_n = \{z = n + \tfrac12\} \,,
\end{equation}
be planes forming a cubic lattice in $\R^3$, where $l$, $m$, $n$ run
over all integers.  Introduce a stack of D4-branes filling $\R^3$ and
place an NS5-brane on each of the planes \eqref{eq:XYZ}.  These branes
also extend in additional directions in ten dimensions, especially a
two-dimensional Minkowski subspace (on which the brane box model
emerges).

The brane box model is a two-dimensional $\CN = (0,2)$ supersymmetric
gauge theory that describes the low-energy dynamics of D4-branes diced
into cubes by NS5-branes as above.  The gauge group of the brane box
model is the product
\begin{equation}
  \prod_{(l,m,n) \in \Z^3} \U(N)_{(l,m,n)} \,,
\end{equation}
where $N$ is the number of D4-branes.  The $\U(N)$ factor
$\U(N)_{(l,m,n)}$ arises from the piece of the stack of $N$ D4-branes
that contains the point $(l,m,n)$.

The brane box model has a $\U(1)$ global symmetry $\U(1)_R$, called an
\emph{R-symmetry}, under which the two supercharges of $\CN = (0,2)$
supersymmetry have charge $+1$ and $-1$.  The definition of $\U(1)_R$
is not unique since the R-charge $R$ generating $\U(1)_R$ can be
shifted by a linear combination of generators of $\U(1)$ flavor
symmetries.  For example, there are $\U(1)$ flavor symmetries
$\U(1)_{X_l}$, $\U(1)_{Y_m}$, $\U(1)_{Z_n}$ associated with $\sfX_l$,
$\sfY_m$, $\sfZ_n$, respectively, so we can assign real numbers
$\xi_l$, $\eta_m$, $\zeta_n$ to $\sfX_l$, $\sfY_m$, $\sfZ_n$ and set
\begin{equation}
  \label{eq:BBM-R}
  R
  =
  R_0
  + \sum_l \xi_l X_l
  + \sum_m \eta_m Y_m
  + \sum_n \zeta_n Z_n
  \,.
\end{equation}
Here $R_0$ is the old R-charge and $X_l$, $Y_m$, $Z_n$ are generators
of $\U(1)_{X_l}$, $\U(1)_{Y_m}$, $\U(1)_{Z_n}$.  The new R-charge
generates a $\U(1)$ global symmetry that transforms the supercharges
in the same way as before and hence still deserves to be called an
R-symmetry.  We will see that parameters assigned to planes, similar
to those just introduced, make an appearance as spectral parameters in
the tetrahedron duality.

In order to describe the matter content of the brane box model, let us
introduce a couple of notations.  For $a$, $b \in \Z^3$, we write
\begin{equation}
  a \xrightarrow{r} b
\end{equation}
and
\begin{equation}
  a \xdashrightarrow{r} b
\end{equation}
to denote, respectively, a chiral multiplet and a Fermi multiplet with
$R = r$ in the bifundamental representation
$(\overline{\mathbf{N}},\mathbf{N})$ of $\U(N)_a \times \U(N)_b$.
There is an equivalence
\begin{equation}
  \label{eq:a2b=b2a}
  a \xdashrightarrow{r} b
  \cong b \xdashrightarrow{-r} a \,,
\end{equation}
which is a general property of $\CN = (0,2)$ supersymmetric field
theories.

For the choice \eqref{eq:BBM-R} of the R-charge, with $R_0 = \frac12$
for chiral multiplets and $0$ for Fermi multiplets, the list of matter
multiplets of the brane box model is the following:
\begin{align}
  \label{eq:BBM-m-start}
  (l,m,n) &\xrightarrow{\frac12+\xi_l} (l+1,m,n) \,,
  \\
  (l,m,n) &\xrightarrow{\frac12+\eta_m} (l,m+1,n) \,,
  \\
  (l,m,n) &\xrightarrow{\frac12+\zeta_n} (l,m,n+1) \,,
  \\
  (l,m,n) &\xleftarrow{\frac12-\xi_l-\eta_m-\zeta_n} (l+1,m+1,n+1) \,,
  \\
  (l,m,n) &\xdashrightarrow{\eta_m+\zeta_n} (l,m+1,n+1) \,,
  \\
  (l,m,n) &\xdashrightarrow{\xi_l+\zeta_n} (l+1,m,n+1) \,,
  \\
  \label{eq:BBM-m-end}
  (l,m,n) &\xdashrightarrow{\xi_l+\eta_m} (l+1,m+1,n) \,.
\end{align}
A matter multiplet has charge $\pm 1$ under $\U(1)_W$, $W = X_l$,
$Y_m$, $Z_n$, if the arrow intersects the corresponding plane
$\mathsf{W}$.  The sign of the charge is correlated with the direction
of intersection.

The matter fields interact with one another via coupling with a vector
multiplet for the gauge group and through a superpotential.  There is
a cubic term in the superpotential for each triangle consisting of two
solid arrows and one dashed arrow with the sum of R-charges equal to
one.  The superpotential is important data of the brane box model, but
we will only be interested in the infrared effective theory in which
all superpotential terms allowed by the symmetries of the model are
expected to be generated.  Thus, in our discussions the ultraviolet
superpotential merely plays the role of a constraint for the possible
charge assignments for $\U(1)_R$ and flavor symmetries, and we will
suppress it in what follows.

Reflecting the lattice structure of the string theory construction,
the brane box model has an interpretation as a spin model on a cubic
lattice, as described in section~\ref{sec:intro}.  For a fixed
$(l,m,n) \in \Z^3$, the matter multiplets
\eqref{eq:BBM-m-start}--\eqref{eq:BBM-m-end} and the eight $\U(N)$
groups connected by them comprise the natural unit cell shown in
Figure~\ref{fig:NUC}.

\subsection{Cancellation of abelian gauge anomaly}

The matter content of the brane box model described above leads to
anomaly in the abelian part of the gauge symmetry.  Let us explain one
way of cancelling this anomaly.  See e.g.~\cites{Quigley:2011pv,
  Blaszczyk:2011ib, Schafer-Nameki:2016cfr} for detailed discussions.

Let $\Sigma$ be the spacetime of the brane box model and $\U(1)_a$ be
the overall $\U(1)$ subgroup of $\U(N)_a$.  Under the gauge
transformation by
\begin{equation}
  \label{eq:AGT}
  (e^{\iu\Lambda_a})_{a \in \Z^3}
  \colon \Sigma \to \prod_{a \in \Z^3} \U(1)_a \,,
\end{equation}
the path integral measure is multiplied by the phase factor
\begin{equation}
  \exp\biggl(
  \iu\sum_{a, b \in \Z^3} \frac{\CA^{ab}}{4\pi} \int_\Sigma \Lambda_a F_b
  \biggr) \,,
\end{equation}
where $F_b$ is the field strength of the gauge field for $\U(1)_b$ and
\begin{equation}
  \CA^{ab}
  =
  \begin{cases}
    2 & a = b \,,
    \\
    -1 & \text{$a \to b$ or $b \to a$} \,,
    \\
    1 & \text{$a \dashrightarrow b$ or $b \dashrightarrow a$} \,,
    \\
    0 & \text{otherwise} \,.
  \end{cases}
\end{equation}

To cancel this phase factor, we introduce a chiral multiplet $\Phi_a$
that is valued in $\C/2\pi\iu\Z$ and transforms under the gauge
transformation \eqref{eq:AGT} as
\begin{equation}
  \Phi_a \to \Phi_a + \iu\Lambda_a \,.
\end{equation}
Then, the gauge transformation of
\begin{equation}
  \label{eq:ACA}
  -\sum_{a,b \in \Z^3} \frac{\CA^{ab}}{4\pi} \int \Im{\phi_a} F_b
\end{equation}
cancels the anomalous variation of the path integral measure, where
$\phi_a$ is the scalar component of $\Phi_a$.  We add the
supersymmetric completion of this term to the action functional of the
brane box model.  Since each nonzero summand is associated with a spin
site (if $a = b$) or an arrow (if $a \neq b$), the added term can be
given a lattice model interpretation.

We expect that the fields $\Im\phi_a$ descend from the Ramond--Ramond
$3$-form field $C^{(3)}$ in type IIA string theory.  For a stack of
D4-branes carrying a gauge field with field strength $F$, the
worldvolume action is given by $\int_{\R^{4,1}} C^{(3)} \wedge \Tr F$.

\subsection{The oblique unit cell and the tetrahedron duality}

The choice of R-charge \eqref{eq:BBM-R} is adapted to the set of
planes $\{\sfX_l, \sfY_m, \sfZ_n\}_{l,m,n \in \Z}$ in the sense that
its parametrization is generated by the flavor symmetries associated
with these planes.  The R-charge assignment relevant for the
tetrahedron duality is adapted to another set of planes
$\{\sfX'_l, \sfY'_m, \sfZ'_n\}_{l,m,n \in \Z}$.

Let $(x',y',z')$ be the coordinates of $\R^3$ related to $(x,y,z)$ by
\begin{equation}
  x' = x \,,
  \qquad
  y' = -y+z \,,
  \qquad
  z' = z \,.
\end{equation}
Then, this set of planes is defined by
\begin{equation}
  \label{eq:X'Y'Z'}
  \sfX'_l = \{x' = l + \tfrac12\} \,,
  \quad
  \sfY'_m = \{y' = m + \tfrac12\} \,,
  \quad
  \sfZ'_n = \{z' = n + \tfrac12\} \,.
\end{equation}
Associated with $\sfX'_l$, $\sfY'_m$, $\sfZ'_n$ there are $\U(1)$
flavor symmetries $\U(1)_{X'_l}$, $\U(1)_{Y'_m}$, $\U(1)_{Z'_n}$,
generated by charges $X'_l$, $Y'_m$, $Z'_n$.  We choose the R-charge
to be
\begin{equation}
  \label{eq:OUC-R}
  R
  =
  R_0
  + \sum_l \Bigl(s_l + \frac12\Bigr) X'_l
  + \sum_m \Bigl(t_m - \frac12\Bigr)  Y'_m
  - \sum_n u_n Z'_n
  \,.
\end{equation}
In terms of the new coordinates and R-charge, the matter content of
the brane box model is
\begin{align}
  \label{eq:BBM-m2-start}
  (l,m,n) &\xrightarrow{1 + s_l} (l+1,m,n) \,,
  \\
  (l+1,m+1,n+1) &\xrightarrow{1 - t_m} (l+1,m,n+1) \,,
  \\
  (l,m,n) &\xrightarrow{t_m - u_n} (l,m+1,n+1) \,,
  \\
  (l,m,n) &\xleftarrow{-s_l + u_n} (l+1,m,n+1) \,,
  \\
  (l+1,m,n) &\xdashrightarrow{-u_n} (l+1,m,n+1) \,,
  \\
  (l,m,n) &\xdashrightarrow{s_l + t_m - u_n} (l+1,m+1,n+1) \,,
  \\
  \label{eq:BBM-m2-end}
  (l,m+1,n+1) &\xdashrightarrow{1+ s_l - t_m} (l+1,m,n+1) \,.
\end{align}

\begin{figure}
  \centering
  \tdplotsetmaincoords{-75}{-35}
  \begin{tikzpicture}[tdplot_main_coords, scale=1.5, rotate=0]
    \begin{scope}[shift={(-1.5,0.5,0)}, scale=0.5]
      \draw[->, shorten >=5pt] (0,0,0) to (1,0,0) node[font=\footnotesize] {$x'$};
      \draw[->, shorten >=5pt] (0,0,0) to (0,1,0) node[font=\footnotesize] {$y'$};
      \draw[->, shorten >=5pt] (0,0,0) to (0,0,1) node[font=\scriptsize] {$z'$};
    \end{scope}

    \node[gnode] (a) at (0,0,0) {$a$};
    \node[gnode] (e) at (1,0,0) {$e$};
    \node[gnode] (f) at (0,1,0) {$f$};
    \node[gnode] (g) at (0,0,1) {$g$};
    \node[gnode] (b) at (0,1,1) {$b$};
    \node[gnode] (c) at (1,0,1) {$c$};
    \node[gnode] (d) at (1,1,0) {$d$};
    \node[gnode] (h) at (1,1,1) {$h$};

    \draw[f->] (a) -- (h);
    \draw[f->] (b) -- (c);
    \draw[f->] (e) -- (c);
    \draw[c->] (a) -- (e);
    \draw[c->] (a) -- (b);
    \draw[c->] (c) -- (a);
    \draw[c->] (h) -- (c);

    \draw (a) -- (f);
    \draw (a) -- (g);
    \draw (e) -- (d);
    \draw (f) -- (d);
    \draw (f) -- (b);
    \draw (g) -- (b);
    \draw (g) -- (c);
    \draw[densely dotted] (b) -- (h);
    \draw[densely dotted] (d) -- (h);
  \end{tikzpicture}

  \caption{The oblique unit cell of the brane box model.}
  \label{fig:cell}
\end{figure}
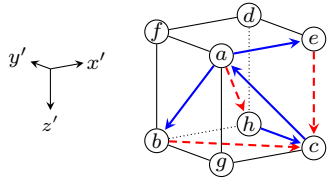

The above list of matter multiplets defines the oblique unit cell
of the brane box model:%
\footnote{The theory defined by the first row of the oblique unit cell
  \eqref{eq:BBM-OUC} also satisfies the tetrahedron duality, but the
  corresponding lattice is the disjoint union of infinitely many
  copies of a two-dimensional lattice.}
\begin{equation}
  \label{eq:BBM-OUC}
  \begin{split}
    (a|e,f,g|b,c,d|h; s, t, u)
    =
    \left(
      \begin{split}
        &
        (a \xrightarrow{t - u} b)
        (b \xdashrightarrow{1 + s - t} c)
        (c \xrightarrow{-s + u} a)
        \\
        &
        (a \xdashrightarrow{s + t - u} h)
        (h \xrightarrow{1 - t} c)
        (a \xrightarrow{1 + s} e)
        (e \xdashrightarrow{-u} c)
      \end{split}
    \right)
    \,.
  \end{split}
\end{equation}
See Figure~\ref{fig:cell}.  By the right-hand side we mean an
$\CN = (0,2)$ supersymmetric field theory consisting of four chiral
multiplets and three Fermi multiplets, each of which transforms in a
bifundamental representation under two of eight $\U(N)$ flavor
symmetry groups $\U(N)_a$, $\U(N)_b$, \dots, $\U(N)_h$.  We emphasize
that these $\U(N)$ groups are not gauged; only when multiple copies of
the oblique unit cell are combined, the $\U(N)$ groups at the
overlapping vertices are gauged.  The oblique unit cell also has a
cubic superpotential, but otherwise its matter fields do not interact
with one another.

Now we argue that the oblique unit cell satisfies the tetrahedron
duality.

Our argument relies on two infrared equivalences.  One is the $(0,2)$
triality, as already highlighted in section~\ref{sec:intro}.  The other is
\begin{equation}
  \label{eq:a2bb2a=1}
  (a \xrightarrow{r} b) (b \xdashrightarrow{1-r} a)
  \simeq
  1
  \,.
\end{equation}
Here $1$ means the trivial multiplet (that is to say, no fields at
all).  This relation holds because in any theory that contains the two
multiplets on the left-hand side, a quadratic term given by their
product can be generated in the superpotential.  Such a superpotential
term makes the pair of multiplets massive and disappear at low
energies.

Plugging the definition \eqref{eq:BBM-OUC} of the oblique unit
cell into the tetrahedron duality~\eqref{eq:TD}, we find that the
left-hand side is given by
\begin{equation}
  \gauge_d
  \left(
  \begin{aligned}
    &
    \underline{(a_4 \xrightarrow{s_3 - s_2} b_1)}_1 \,
    \underline{(b_1 \xdashrightarrow{1 + s_1 - s_3} b_3)}_2 \,
    \underline{(b_3 \xrightarrow{-s_1 + s_2} a_4)}_3
    \\
    &(a_4 \xdashrightarrow{s_1 + s_3 - s_2} d)
    \underline{(d \xrightarrow{1 - s_3} b_3)}_4 \,
    \underline{(a_4 \xrightarrow{1 + s_1} c_2)}_5 \,
    \underline{(c_2 \xdashrightarrow{-s_2} b_3)}_6
    \\
    &(c_1 \xrightarrow{s_4 - s_2} c_4)
    (c_4 \xdashrightarrow{1 + s_1 - s_4} d)
    (d \xrightarrow{-s_1 + s_2} c_1)
    \\
    &(c_1 \xdashrightarrow{s_1 + s_4 - s_2} b_4)
    (b_4 \xrightarrow{1 - s_4} d)
    \underline{(c_1 \xrightarrow{1 + s_1} b_2)}_7 \,
    \underline{(b_2 \xdashrightarrow{-s_2} d)}_8
    \\
    &
    \underline{(b_1 \xrightarrow{s_4 - s_3} a_2)}_9 \,
    \underline{(a_2 \xdashrightarrow{1 + s_1 - s_4} b_3)}_{10} \,
    \underline{(b_3 \xrightarrow{-s_1 + s_3} b_1)}_2
    \\
    &(b_1 \xdashrightarrow{s_1 + s_4 - s_3} c_5)
    \underline{(c_5 \xrightarrow{1 - s_4} b_3)}_{11}
    (b_1 \xrightarrow{1 + s_1} d)
    \underline{(d \xdashrightarrow{-s_3} b_3)}_4
    \\
    &(d \xrightarrow{s_4 - s_3} c_5)
    (c_5 \xdashrightarrow{1 + s_2 - s_4} c_2)
    (c_2 \xrightarrow{-s_2 + s_3} d)
    \\
    &(d \xdashrightarrow{s_2 + s_4 - s_3} a_1)
    \underline{(a_1 \xrightarrow{1 - s_4} c_2)}_{12} \,
    \underline{(d \xrightarrow{1 + s_2} b_2)}_8 \,
    \underline{(b_2 \xdashrightarrow{-s_3} c_2)}_{13}
  \end{aligned}
\right)
\,,
\end{equation}
whereas the right-hand side is
\begin{equation}
  \gauge_d
  \left(
  \begin{aligned}
    &
    \underline{(b_1 \xrightarrow{s_4 - s_3} a_2)}_9 \,
    \underline{(a_2 \xdashrightarrow{1 + s_2 - s_4} a_4)}_{14} \,
    \underline{(a_4 \xrightarrow{-s_2 + s_3} b_1)}_1
    \\
    &(b_1 \xdashrightarrow{s_2 + s_4 - s_3} d)
    (d \xrightarrow{1 - s_4} a_4)
    (b_1 \xrightarrow{1 + s_2} c_1)
    (c_1 \xdashrightarrow{-s_3} a_4)
    \\
    &(c_1 \xrightarrow{s_4 - s_3} d)
    (d \xdashrightarrow{1 + s_1 - s_4} c_2)
    (c_2 \xrightarrow{-s_1 + s_3} c_1)
    \\
    &(c_1 \xdashrightarrow{s_1 + s_4 - s_3} a_1)
    \underline{(a_1 \xrightarrow{1 - s_4} c_2)}_{12} \,
    \underline{(c_1 \xrightarrow{1 + s_1} b_2)}_7 \,
    \underline{(b_2 \xdashrightarrow{-s_3} c_2)}_{13}
    \\
    &
    \underline{(a_4 \xrightarrow{s_4 - s_2} a_2)}_{14} \,
    \underline{(a_2 \xdashrightarrow{1 + s_1 - s_4} b_3)}_{10} \,
    \underline{(b_3 \xrightarrow{-s_1 + s_2} a_4)}_3
    \\
    &(a_4 \xdashrightarrow{s_1 + s_4 - s_2} c_5)
    \underline{(c_5 \xrightarrow{1 - s_4} b_3)}_{11} \,
    \underline{(a_4 \xrightarrow{1 + s_1} c_2)}_5 \,
    \underline{(c_2 \xdashrightarrow{-s_2} b_3)}_6
    \\
    &(d \xrightarrow{s_3 - s_2} c_4)
    (c_4 \xdashrightarrow{1 + s_1 - s_3} c_5)
    (c_5 \xrightarrow{-s_1 + s_2} d)
    \\
    &
    (d \xdashrightarrow{s_1 + s_3 - s_2} b_4)
    (b_4 \xrightarrow{1 - s_3} c_5)
    (d \xrightarrow{1 + s_1} a_1)
    (a_1 \xdashrightarrow{-s_2} c_5)
  \end{aligned}
\right)
\,.
\end{equation}

Among the 56 arrows that appear in the tetrahedron duality, 14 pairs
(underlined and numbered above) cancel against each other by virtue of
the equivalences \eqref{eq:a2b=b2a} and \eqref{eq:a2bb2a=1}.  After
the cancellations we have
\begin{equation}
  \gauge_d
  \left(
  \begin{aligned}
    &
    (d \xrightarrow{-s_1 + s_2} c_1)
    (d \xrightarrow{s_4 - s_3} c_5)
    \\
    &
    (d \xdashrightarrow{s_2 + s_4 - s_3} a_1)
    \underline{(a_4 \xdashrightarrow{s_1 + s_3 - s_2} d)} \,
    \underline{(c_4 \xdashrightarrow{1 + s_1 - s_4} d)}
    \\
    &
    (b_1 \xrightarrow{1 + s_1} d)
    (b_4 \xrightarrow{1 - s_4} d)
    (c_2 \xrightarrow{-s_2 + s_3} d)
    \\
    &
    (c_1 \xdashrightarrow{s_1 + s_4 - s_2} b_4)
    (c_1 \xrightarrow{s_4 - s_2} c_4)
    \underline{(b_1 \xdashrightarrow{s_1 + s_4 - s_3} c_5)}
    (c_5 \xdashrightarrow{1 + s_2 - s_4} c_2)
  \end{aligned}
\right)
\end{equation}
and
\begin{equation}
  \gauge_d
  \left(
  \begin{aligned}
    &
    (c_1 \xrightarrow{s_4 - s_3} d)
    (c_5 \xrightarrow{-s_1 + s_2} d)
    \\
    &
    (d \xrightarrow{1 + s_1} a_1)
    (d \xrightarrow{1 - s_4} a_4)
    (d \xrightarrow{s_3 - s_2} c_4)
    \\
    &
    \underline{(b_1 \xdashrightarrow{s_2 + s_4 - s_3} d)}
    (d \xdashrightarrow{s_1 + s_3 - s_2} b_4)
    (d \xdashrightarrow{1 + s_1 - s_4} c_2)
    \\
    &
    (b_1 \xrightarrow{1 + s_2} c_1)
    \underline{(c_1 \xdashrightarrow{-s_3} a_4)}
    (c_2 \xrightarrow{-s_1 + s_3} c_1)
    \underline{(c_1 \xdashrightarrow{s_1 + s_4 - s_3} a_1)}
    \\
    &
    (a_4 \xdashrightarrow{s_1 + s_4 - s_2} c_5)
    (c_4 \xdashrightarrow{1 + s_1 - s_3} c_5)
    (b_4 \xrightarrow{1 - s_3} c_5)
    (a_1 \xdashrightarrow{-s_2} c_5)
  \end{aligned}
\right)
\end{equation}
for the left-hand side and the right-hand side.  In this step we have
used the equivalence \eqref{eq:a2b=b2a} once (when cancelling pair
\#4).

Further applying the equivalence \eqref{eq:a2b=b2a} to the six
underlined arrows and moving some arrows across the two sides using
the equivalence \eqref{eq:a2bb2a=1}, we arrive at the infrared duality
\begin{multline}
  \label{eq:TD-simple}
  \gauge_d
  \left(
  \begin{aligned}
    &
    (d \xrightarrow{-s_1 + s_2} c_1)
    (d \xrightarrow{s_4 - s_3} c_5)
    \\
    &
    (d \xdashrightarrow{s_2 + s_4 - s_3} a_1)
    (d \xdashrightarrow{-s_1 - s_3 + s_2} a_4)
    (d \xdashrightarrow{-1 - s_1 + s_4} c_4)
    \\
    &
    (b_1 \xrightarrow{1 + s_1} d)
    (b_4 \xrightarrow{1 - s_4} d)
    (c_2 \xrightarrow{-s_2 + s_3} d)
    \\
    &
    (c_1 \xdashrightarrow{-s_2} b_1)
    (c_1 \xdashrightarrow{s_1 + s_4 - s_2} b_4)
    (c_1 \xdashrightarrow{1 + s_1 - s_3} c_2)
    \\
    &
    (c_5 \xdashrightarrow{-s_1 - s_4 + s_3} b_1)
    (c_5 \xdashrightarrow{s_3} b_4)
    (c_5 \xdashrightarrow{1 + s_2 - s_4} c_2)
  \end{aligned}
  \right)
  \\
  \simeq
  \gauge_d
  \left(
  \begin{aligned}
    &
    (c_1 \xrightarrow{s_4 - s_3} d)
    (c_5 \xrightarrow{-s_1 + s_2} d)
    \\
    &
    (d \xrightarrow{1 + s_1} a_1)
    (d \xrightarrow{1 - s_4} a_4)
    (d \xrightarrow{s_3 - s_2} c_4)
    \\
    &
    (d \xdashrightarrow{-s_2 - s_4 + s_3} b_1)
    (d \xdashrightarrow{s_1 + s_3 - s_2} b_4)
    (d \xdashrightarrow{1 + s_1 - s_4} c_2)
    \\
    &
    (a_1 \xdashrightarrow{-s_1 - s_4 + s_3} c_1)
    (a_4 \xdashrightarrow{s_3} c_1)
    (c_4 \xdashrightarrow{1 - s_4 + s_2} c_1)
    \\
    &
    (a_1 \xdashrightarrow{-s_2} c_5)
    (a_4 \xdashrightarrow{s_1 + s_4 - s_2} c_5)
    (c_4 \xdashrightarrow{1 + s_1 - s_3} c_5)
  \end{aligned}
  \right)
  \,.
\end{multline}

The two sides of this relation are two of a triplet of $\CN = (0,2)$
supersymmetric gauge theories related by the $(0,2)$ triality.
Indeed, they are the theories $T_{3N,2N,3N}$ and $T_{2N,3N,3N}$ in the
notation of Appendix~\ref{sec:triality}, except that a different mechanism
is employed for the cancellation of abelian gauge anomaly.  Therefore,
the tetrahedron duality holds under the assumption that the $(0,2)$
triality is not broken by the anomaly cancellation mechanism of the
brane box model.

\subsection{Spectral parameters and boundary conditions}

An elliptic genus of the oblique unit cell (or any unit cell for that
matter) defines a Boltzmann weight for a lattice model whose partition
function is equal to the elliptic genus of the brane box model.  In
order to make the partition function finite, one usually makes the
lattice finite by some means.  For example, we can impose periodic
boundary conditions on the lattice, placing it in a $3$-torus $\T^3$
rather than $\R^3$.  Let us take the size of the resulting lattice to
be $L \times M \times N$.

The partition function of the lattice model is a function of the
spectral parameters, which are assigned to the planes
\eqref{eq:X'Y'Z'} and appear in the definition of the R-charge
\eqref{eq:OUC-R}.  Although the total number of the spectral
parameters is $L+M+N$, the partition function actually depends only on
three linear combinations of them because of gauge symmetry.

For any oriented closed surface $\mathsf{S}$ in $\T^3$ that does not
intersect with the integral points, there is an associated $\U(1)$
symmetry $\U(1)_S$ in the brane box model, defined in a similar manner
to $\U(1)_{X'_l}$, $\U(1)_{Y'_m}$, $\U(1)_{Z'_n}$.  If $\mathsf{S}$ is
the boundary of a three-dimensional region $\mathsf{V}$ of $\T^3$,
then $\U(1)_S$ is a gauge symmetry: for
$\mathsf{S} = \partial\mathsf{V}$ with orientation given by the normal
vectors pointing outward,
\begin{equation}
  \U(1)_S = \prod_{(l,m,n) \in \mathsf{V}} \U(1)_{(l,m,n)} \,.
\end{equation}
Thus, the number of independent $\U(1)$ flavor symmetries is given by
the rank of the second homology group of $\T^3$, which is $3$.
Without loss of generality, we can set all spectral parameters but
those assigned to $\sfX'_0$, $\sfY'_0$, $\sfZ'_0$ to zero.  The three
remaining parameters $s_0$, $t_0$, $u_0$ may be interpreted as
specifying twisted boundary conditions in the $x'$, $y'$, $z'$
directions, respectively.

\section{A solution by the 1D brane box model}
\label{sec:1D-BBM}

Dimensional reduction of the brane box model yields the 1D brane box
model, an $\CN = 2$ supersymmetric gauge theory that has the same
gauge group and matter content as its parent theory (except that the
chiral and Fermi multiplets are, of course, those for $\CN = 2$
supersymmetry in one dimension).  If the dimensional reduction
commutes with the triality, then the oblique unit cell for the 1D
brane box model also satisfies the tetrahedron duality and its Witten
index, computed by a partition function on a circle, is a solution of
the tetrahedron equation.

Crucially, the abelian gauge anomaly disappears after the dimensional
reduction since chiral anomalies are absent in one dimension.  Thus,
we can write down the Witten index by the dimensional reduction of an
integral formula for the relevant elliptic genus \cites{Benini:2013nda,
  Benini:2013xpa}.  The formula shows that the Witten index of the 1D
brane box model is given by the partition function of an Ising-type
spin model with continuous spin variables and pairwise interaction.
Below we describe this lattice model and verify that its Boltzmann
weight indeed solves the tetrahedron equation.

\subsection{The lattice model}
\label{sec:QM-lattice}

The spin model in question is defined on a simple cubic lattice.  At
each point $a \in \Z^3$ a set of $N$ periodic spin variables
\begin{equation}
  a = (a_i)_{i=1}^N \in \T^N
\end{equation}
is placed, where $\T \simeq \R/2\pi\Z$ is the unit circle in the
complex plane.  We will use the same symbol to refer to both a spin
site and the set of spin variables located there.

The interactions between spin variables at neighboring sites come in
two types and are represented by solid arrows and dashed arrows in the
list \eqref{eq:BBM-m2-start}--\eqref{eq:BBM-m2-end}.  The Boltzmann
weights for these interactions are given by%
\footnote{Note that only the non-integer part of $r$ appears in the
  Boltzmann weights.  In the language of the Witten index,
  $e^{4\pi\iu r}$ is the product of the flavor fugacities for
  $U(1)_{X'_l}$, $U(1)_{Y'_m}$, $U(1)_{Z'_n}$.}
\begin{align}
  W(a \xrightarrow{r} b)
  &= \prod_{i,j} -\frac{1}{\phi(e^{4\pi\iu r} b_i/a_j)} \,,
    \label{eq:1DBW-c}
  \\
  W(a \xdashrightarrow{r} b) &= \prod_{i,j} \phi(e^{4\pi\iu r} b_i/a_j) \,,
\end{align}
where
\begin{equation}
  \phi(z) = z^{1/2} - z^{-1/2}
  \,.
\end{equation}
The number $r$ above an arrow is a coupling constant controlling the
interaction.  We take $r \in \R$.

Since $\phi(z) = -\phi(1/z)$, the equivalence \eqref{eq:a2b=b2a}
becomes an equality up to a sign for the Boltzmann weights:
\begin{equation}
  \label{eq:Wab=-Wba}
  W(a \xdashrightarrow{r} b) = (-1)^N W(b \xdashrightarrow{-r} a) \,.
\end{equation}
The cancellation relation \eqref{eq:a2bb2a=1} holds without a sign:
\begin{equation}
  \label{eq:Wa2bWb2a=1}
  W(a \xrightarrow{r} b) W(b \xdashrightarrow{1-r} a) = 1 \,.
\end{equation}

The partition function of the lattice model is the integral over all
spin variables of the product of all local Boltzmann weights.  The
integration over spin variables $a \in \T^N$ is performed with respect
to the measure
\begin{equation}
  \mu(a)
  =
  \frac{1}{N!}
  \prod_i \frac{\rmd a_i}{2\pi\iu a_i}
  \prod_{i \neq j} \phi\biggl(\frac{a_i}{a_j}\biggr) \,.
\end{equation}
The factor $\prod_{i \neq j} \phi(a_i/a_j)$ may be thought as the
Boltzmann weight for self-interaction among $(a_i)_{i=1}^N$.

In performing this integration we encounter poles coming from factors
of the form $W(b \xrightarrow{r} a)$ and factors of the form
$W(a \xrightarrow{r} b)$.  We agree to deform the domain of
integration in such a way that the compact region bounded by the
domain contains all poles from the former but excludes any poles from
the latter.  Alternatively, we may modify the definition of the
Boltzmann weight \eqref{eq:1DBW-c} by multiplying the argument of the
function $\phi$ by $e^{\epsilon}$ and later take the limit
$\epsilon \to +0$.

\subsection{The tetrahedron equation}
\label{sec:QM-TE}

The tetrahedron equation for the oblique unit cell of this lattice
model simplifies to the following equality, obtained by evaluating the
Witten indices of the two sides of the equivalence relation
\eqref{eq:TD-simple}:
\begin{multline}
  \label{eq:TD-simple-1D}
  \int_d \mu(d)
  \left(
  \begin{aligned}
    &
    W(d \xrightarrow{-s_1 + s_2} c_1)
    W(d \xrightarrow{s_4 - s_3} c_5)
    \\ \times
    &
    W(d \xdashrightarrow{s_2 + s_4 - s_3} a_1)
    W(d \xdashrightarrow{-s_1 - s_3 + s_2} a_4)
    W(d \xdashrightarrow{-1 - s_1 + s_4} c_4)
    \\ \times
    &
    W(b_1 \xrightarrow{1 + s_1} d)
    W(b_4 \xrightarrow{1 - s_4} d)
    W(c_2 \xrightarrow{-s_2 + s_3} d)
    \\ \times
    &
    W(c_1 \xdashrightarrow{-s_2} b_1)
    W(c_1 \xdashrightarrow{s_1 + s_4 - s_2} b_4)
    W(c_1 \xdashrightarrow{1 + s_1 - s_3} c_2)
    \\ \times
    &
    W(c_5 \xdashrightarrow{-s_1 - s_4 + s_3} b_1)
    W(c_5 \xdashrightarrow{s_3} b_4)
    W(c_5 \xdashrightarrow{1 + s_2 - s_4} c_2)
  \end{aligned}
  \right)
  \\
  =
  (-1)^N \int_d \mu(d)
  \left(
  \begin{aligned}
    &
    W(c_1 \xrightarrow{s_4 - s_3} d)
    W(c_5 \xrightarrow{-s_1 + s_2} d)
    \\ \times
    &
    W(d \xrightarrow{1 + s_1} a_1)
    W(d \xrightarrow{1 - s_4} a_4)
    W(d \xrightarrow{s_3 - s_2} c_4)
    \\ \times
    &
    W(d \xdashrightarrow{-s_2 - s_4 + s_3} b_1)
    W(d \xdashrightarrow{s_1 + s_3 - s_2} b_4)
    W(d \xdashrightarrow{1 + s_1 - s_4} c_2)
    \\ \times
    &
    W(a_1 \xdashrightarrow{-s_1 - s_4 + s_3} c_1)
    W(a_4 \xdashrightarrow{s_3} c_1)
    W(c_4 \xdashrightarrow{1 - s_4 + s_2} c_1)
    \\ \times
    &
    W(a_1 \xdashrightarrow{-s_2} c_5)
    W(a_4 \xdashrightarrow{s_1 + s_4 - s_2} c_5)
    W(c_4 \xdashrightarrow{1 + s_1 - s_3} c_5)
  \end{aligned}
  \right)
  \,.
\end{multline}
The sign $(-1)^N$ comes from the use of the relation
\eqref{eq:a2b=b2a} seven times in the simplification of the
tetrahedron duality.

Let $N_1$, $N_2$, $N_3$, $n_2$, $n_3$ be nonnegative integers such that
\begin{equation}
  \label{eq:n2n3}
  n_2 = \frac12 (N_1 - N_2 + N_3) \,,
  \qquad
  n_3 = \frac12 (N_1 + N_2 - N_3) \,.
\end{equation}
Then, we have the following equality
\begin{multline}
  \label{eq:1D-triality}
  \frac{(-1)^{n_3 N_2}}{n_3!} \int_\cD
  \prod_{i} \frac{\rmd z_i}{2\pi\iu z_i}
  \prod_{i \neq j}
  \phi\biggl(\frac{z_i}{z_j}\biggr)
  \frac{
    \prod_{i,\gamma} \phi(\tau_\gamma w_\gamma/z_i)
    \prod_{\alpha,\beta} \phi(u_\alpha/\rho_\alpha \sigma_\beta v_\beta)
  }{
    \prod_{i,\alpha} \phi(\rho_\alpha z_i/u_\alpha)
    \prod_{i,\beta} \phi(\sigma_\beta v_\beta/z_i)
  }
  \\
  =
  \frac{(-1)^{n_2(N_2 + N_3)}}{n_2!} \int_{\cD'}
  \prod_{\ib} \frac{\rmd z'_\ib}{2\pi\iu z'_\ib}
  \prod_{\ib \neq \jb}
  \phi\biggl(\frac{z'_\ib}{z'_\jb}\biggr)
  \frac{
    \prod_{\ib,\beta} \phi(\sigma'_\beta v_\beta/z'_\ib)
    \prod_{\gamma,\alpha} \phi(w_\gamma/\tau'_\gamma \rho'_\alpha u_\alpha)
  }{
    \prod_{\ib,\gamma} \phi(\tau'_\gamma z'_\ib/w_\gamma)
    \prod_{\ib,\alpha} \phi(\rho'_\alpha u_\alpha/z'_\ib)
  }
  \,.
\end{multline}
Here
\begin{equation}
  (u_\alpha, \rho_\alpha)_{\alpha=1}^{N_1} \in \T^{2N_1} \,,
  \quad
  (v_\beta, \sigma_\beta)_{\beta=1}^{N_2} \in \T^{2N_2} \,,
  \quad
  (w_\gamma, \tau_\gamma)_{\gamma=1}^{N_3} \in \T^{2N_3} \,,
\end{equation}
are arbitrary parameters and
\begin{equation}
  \label{eq:sigma'}
  \rho'_\alpha = e^{4\pi\iu \Delta} \rho_\alpha^{-1} \,,
  \quad
  \sigma'_\beta = e^{4\pi\iu(-1 + \Delta)} \sigma_\beta \,,
  \quad
  \tau'_\gamma = e^{4\pi\iu(1 - \Delta)} \tau_\alpha^{-1} \,,
\end{equation}
with $\Delta \in \R$ arbitrary and fixed.  For each variable $z_i$,
$i = 1$, \dots, $n_3$, the integrand of the left-hand side is a
meromorphic differential on $\mathbb{CP}^1$ which have no poles at
$z_i = 0$ and $\infty$; all poles come from the factors
$\prod_{i,\alpha} \phi(\rho_\alpha z_i/u_\alpha)^{-1}$ and
$\prod_{i,\beta} \phi(\sigma_\beta v_\beta/z_i)^{-1}$.  The domain
$\cD$ bounds a region that contains $0$ and all poles from the former
factor but no poles from the latter.  Similarly, for each variable
$z'_\ib$, $\ib = 1$, \dots, $n_2$, the integrand of the right-hand
side is a meromorphic differential on $\mathbb{CP}^1$ with no poles at
$z'_\ib = 0$ and $\infty$, and $\cD'$ bounds a region that contains
all poles from the factor
$\prod_{i,\gamma} \phi(\tau'_\gamma z'_\ib/w_\gamma)^{-1}$ but no
poles from
$\prod_{\ib,\alpha} \phi(\rho'_\alpha u_\alpha/z'_\ib)^{-1}$.

Equation \eqref{eq:TD-simple-1D} is a special case of
\eqref{eq:1D-triality}, hence the oblique unit cell is a solution of
the tetrahedron equation.  Indeed, if we take $N_1 = 2N$,
$N_2 = N_3 = 3N$ (for which $n_2 = n_3 = N$),
\begin{alignat}{2}
  (u_\alpha)_{\alpha=1}^{N_1} &= (c_1, c_5) \,,
  & \quad
  \rho_\alpha
  &=
  \begin{cases}
    e^{4\pi\iu(s_4 - s_3)} & (1 \leq \alpha \leq N) \,,
    \\
    e^{4\pi\iu(-s_1 + s_2)} & (N + 1 \leq \alpha \leq 2N) \,,
  \end{cases}
  \\
  (v_\beta)_{\beta=1}^{N_2} &= (a_1, a_4, c_4) \,,
  & \quad
  \sigma_\beta
  &= 
  \begin{cases}
    e^{4\pi\iu(1 + s_1)} & (1 \leq \alpha \leq N) \,,
    \\
    e^{4\pi\iu(1 - s_4)} & (N + 1 \leq \alpha \leq 2N) \,,
    \\
    e^{4\pi\iu(s_3 - s_2)} & (2N + 1 \leq \alpha \leq 3N) \,,
  \end{cases}
  \\
  (w_\gamma)_{\gamma=1}^{N_3} &= (b_1, b_4, c_2) \,,
  & \quad
  \tau_\gamma
  &= 
  \begin{cases}
    e^{4\pi\iu(-s_2 - s_4 + s_3)} & (1 \leq \alpha \leq N) \,,
    \\
    e^{4\pi\iu(s_1 + s_3 - s_2)} & (N + 1 \leq \alpha \leq 2N) \,,
    \\
    e^{4\pi\iu(1 + s_1 - s_4)} & (2N + 1 \leq \alpha \leq 3N) \,,
  \end{cases}
\end{alignat}
and
\begin{equation}
  \Delta = s_4 - s_3 - s_1 + s_2 \,,
\end{equation}
then the left-hand side and the right-hand of \eqref{eq:1D-triality}
become the right-hand side and the left-hand side of
\eqref{eq:TD-simple-1D}, respectively.

Equation \eqref{eq:1D-triality} can be demonstrated in a way similar
to the proof of the equality between the elliptic genera of theories
related by the $(0,2)$ triality, treated in detail in
Appendix~\ref{sec:triality}.  Using
\begin{equation}
  \Res_{z = w} \frac{1}{z \phi(z/w)}
  = 1
  \,,
\end{equation}
we find that the left-hand side evaluates to
\begin{equation}
  \sum_{\substack{\FS \subset \{1, \dotsc, N_1\} \\ |\FS| = n_3}}
  \frac{
    \prod_{i \in \FS} \prod_\gamma \phi(\tau_\gamma \rho_i w_\gamma/u_i)
    \prod_{\ib \in \FSb} \prod_\beta \phi(u_\ib/\rho_\ib \sigma_\beta v_\beta)
  }{
    \prod_{i \in \FS} \prod_{\ib \in \FSb}
    \phi(\rho_\ib u_i/\rho_i u_\ib)
  }
  \,,
\end{equation}
where $\FSb = \{1, \dotsc, N_1\} \setminus \FS$.  For the right-hand
side, we can multiply it by $(-1)^{n_2}$ and sum over the residues at
the poles from the factor
$\prod_{\ib,\alpha} \phi(\rho'_\alpha u_\alpha/z'_\ib)^{-1}$ since
the sum of the resides of a meromorphic differential on any compact
Riemann surface is zero.  The sign factor $(-1)^{n_2}$ just introduced
is cancelled by the same factor coming from the residues, and we get
\begin{equation}
  \sum_{\substack{\FSb \subset \{1, \dotsc, N_1\} \\ |\FSb| = n_2}}
  \frac{
    \prod_{\ib \in \FSb} \prod_\beta \phi(\rho'_\ib u_\ib/\sigma'_\beta v_\beta)
    \prod_{i \in \FS} \prod_\gamma \phi(w_\gamma/\tau'_\gamma \rho'_i u_i)
  }{
    \prod_{i \in \FS} \prod_{\ib \in \FSb}
    \phi(\rho'_i u_i/\rho'_\ib u_\ib)
  }
  \,,
\end{equation}
with $\FS = \{1, \dotsc, N_1\} \setminus \FSb$.  Looking at the
relation \eqref{eq:sigma'} we see that the above two sums are equal.

\section{A solution by the \texorpdfstring{$(0,1)$}{(0,1)} brane box
  model}
\label{sec:(0,1)-BBM}

$\CN = (0,1)$ supersymmetric gauge theories with special orthogonal
gauge groups are also known to exhibit triality \cite{Gukov:2019lzi}.
Since this triality has the same combinatorial structure as its
$(0,2)$ counterpart, we obtain another solution of the tetrahedron
duality by mere reinterpretation of notations that turns the brane box
model into the relevant $\CN = (0,1)$ supersymmetric gauge theory,
which we call the $(0,1)$ brane box model.

For the $(0,1)$ brane box model, each point $a \in \Z^3$ represents an
$\SO(2N)$ gauge group $\SO(2N)_a$, accompanied by a Fermi multiplet in
the symmetric representation of $\SO(2N)_a$.  A solid arrow $a \to b$
and a dashed arrow $a \xdashrightarrow{} b$ are a chiral multiplet and
a Fermi multiplet in the bivector representation of
$\SO(2N)_a \times \SO(2N)_b$, respectively.%
\footnote{From the point of view of field theory there is nothing that
  distinguishes $a \to b$ from $b \to a$ or $a \xdashrightarrow{} b$
  from $b \xdashrightarrow{} a$.  Nevertheless, we will keep the
  orientations of arrows because somehow they carry essential
  information with regard to the integral formula for the elliptic
  genus of the $(0,1)$ brane box model.}
All of these multiplets are those for $\CN = (0,1)$ supersymmetry.  In
particular, there is no R-symmetry and hence no number (R-charge)
assigned to arrows.  The $(0,1)$ brane box model is free of gauge
anomaly without the need of introducing additional matter multiplets.

Although no formula is known for the elliptic genera of $\CN = (0,1)$
supersymmetric gauge theories in general, in the case relevant to the
$(0,1)$ triality the elliptic genus in the Ramond--Ramond sector has
been computed in the low-energy effective sigma model description
\cite{Gukov:2019lzi}.  It turns out that the result can be expressed
as an integral akin to integral formulas for the elliptic genera of
$\CN = (0,2)$ supersymmetric gauge theories.  Based on this
observation we propose that the elliptic genus of the $(0,1)$ brane
box model is equal to the partition function of a lattice model
defined by a solution of the tetrahedron equation, which we now
describe.

\subsection{The lattice model}

Fix a complex number $q$ with $|q| < 1$.  Let
\begin{equation}
  \eta = q^{1/24} \prod_{n=1}^\infty (1 - q^n)
\end{equation}
be the Dedekind eta function and
\begin{equation}
  \thetah(z)
  =
  \iu \frac{\theta_1(z)}{\eta}
  =
  q^{1/12} (z^{1/2} - z^{-1/2})
  \prod_{n=1}^\infty (1 - zq^n)(1 - z^{-1} q^n) \,,
\end{equation}
where $\theta_1$ is the Jacobi theta function.  Note that
$q^{-1/12} \thetah(z) \to \phi(z)$ as $q \to 0$; this limit
corresponds to dimensional reduction.  We have
$\thetah(z) = -\thetah(1/z)$ and
\begin{equation}
  \thetah(qz) = -q^{-1/2} z^{-1} \thetah(z) \,.
\end{equation}

The lattice model arising from the elliptic genus of the $(0,1)$ brane
box model is much like the one for the 1D brane box model described
in section~\ref{sec:QM-lattice}.  A set of continuous spin variables
$a = (a_i)_{i=1}^N$ resides at $a \in \Z^3$.  They interact according
to the list \eqref{eq:BBM-m2-start}--\eqref{eq:BBM-m2-end} (with the
numbers above the arrows ignored).  The Boltzmann weights for these
interactions are
\begin{align}
  W(a \to b)
  &= \prod_{i,j} -\frac{1}{\thetah(b_i a_j^{\pm1})} \,,
  \\
  W(a \xdashrightarrow{} b) &= \prod_{i,j} \thetah(b_i a_j^{\pm1})
\end{align}
and satisfy the relations \eqref{eq:Wab=-Wba} and
\eqref{eq:Wa2bWb2a=1}.  Here we have introduced the shorthand notation
$\thetah(b_i a_j^{\pm 1}) = \thetah(b_i a_j) \thetah(b_i a_j^{-1})$.
The integration over spin variables $a$ is performed with respect to
the measure
\begin{equation}
  \mu(a)
  =
  \frac{1}{2^{N - 1} N!}
  \prod_i \frac{\rmd a_i}{2\pi\iu a_i} \eta^2 \thetah(a_i^2)
  \prod_{i \neq j} \thetah(a_i a_j^{\pm1})
  \,.
\end{equation}
The numerical factor $2^{N-1} N!$ is the order of the Weyl group of
$\SO(2N)$.

The choice of the domain of integration is slightly more complicated
here than the case of the 1D brane box model.  As a function of $a_i$,
$W(b \to a)$ has poles at $a_i = q^n b_j^{\pm1}$, $n \in \Z$.  The
integration contour for $a_i$ is such that it encircles all poles of
the form $a_i = b_j^{\pm1}$ in $W(b \to a)$ and excludes all other
poles.  In the 1D limit $q \to 0$, the poles outside the contour are
all sent to $0$ or $\infty$.

\subsection{The tetrahedron equation}

The proof of the tetrahedron equation for the oblique unit cell of
this lattice model is entirely analogous to the case of the 1D brane
box model in section~\ref{sec:QM-TE}.

Let $N_1$, $N_2$, $N_3$, $n_2$, $n_3$ be positive integers satisfying
\eqref{eq:n2n3}.  The tetrahedron equation is a special case of the
following equation:
\begin{multline}
  \label{eq:(0,1)-triality}
  \frac{(-1)^{n_3 N_2}}{2^{n_3 - 1} n_3!}
  \int_\cD
  \prod_i \frac{\rmd z_i}{2\pi\iu z_i}
  \eta^{2} \thetah(z_i^2)
  \prod_{i \neq j} \thetah(z_i z_j^{\pm1})
  \frac{
    \prod_{i,\gamma} \thetah(w_\gamma z_i^{\pm1})
    \prod_{\alpha,\beta} \thetah(u_\alpha v_\beta^{\pm1})
  }{
    \prod_{i,\alpha} \thetah(z_i u_\alpha^{\pm1})
    \prod_{i,\beta} \thetah(v_\beta z_i^{\pm1})
  }
  \\
  =
  \frac{(-1)^{n_2(N_2 + N_3)}}{2^{n_2 - 1} n_2!}
  \int_{\cD'}
  \prod_\ib \frac{\rmd z_\ib'}{2\pi\iu z_i}
  \eta^{2} \thetah(z_\ib'^2)
  \prod_{\ib \neq \jb} \thetah(z_\ib' z_\jb'^{\pm1})
  \frac{
    \prod_{\ib,\beta} \thetah(v_\beta z_\ib'^{\pm1})
    \prod_{\gamma,\alpha} \thetah(w_\gamma u_\alpha^{\pm1})
  }{
    \prod_{\ib,\gamma} \thetah(z'_\ib w_\gamma^{\pm1})
    \prod_{\ib,\alpha} \thetah(u_\alpha z_\ib'^{\pm1})
  }
  \,.
\end{multline}
The domains $\cD$ and $\cD'$ enclose all poles where
$\{z_i\}_{i=1}^{n_3} \subset \{u_\alpha^{\pm1}\}_{\alpha=1}^{N_1}$ and
$\{z'_\ib\}_{\ib=1}^{n_2} \subset
\{w_\gamma^{\pm1}\}_{\gamma=1}^{N_3}$, respectively, but no other poles.
The integrand on each side is a meromorphic differential on the
elliptic curve $\C^\times/q\Z$ for each integration variable.

Let $\FS$ be a subset of $\{1, \dotsc, N_1\}$ of order $n_3$ and
consider the set of poles where
$\{z_i\}_{i=1}^{n_3} \subset \{u_i^{\pm1}\}_{i \in \FS}$ in
the integrand on the left-hand side.  The poles where
$z_i = z_j^{\pm1}$ for some $i$, $j$ with $i \neq j$ do not
contribute to the integral because of the factor
$\thetah(z_i z_j^{\pm1}$).  There are $2^{n_3} n_3!$ remaining
poles from this set.  Using
\begin{equation}
  \Res_{z = w} \frac{\eta^2}{z \thetah(z/w)}
  = 1
  \,,
\end{equation}
we find that the sum of their residues is
\begin{equation}
  \label{eq:(0,1)-SOR}
  2
  \frac{
    \prod_{i \in \FS} \prod_\gamma \thetah(w_\gamma u_i^{\pm1})
    \prod_{\ib \in \FSb} \prod_\beta \thetah(u_\ib v_\beta^{\pm1})
  }{
    \prod_{i \in \FS} \prod_{\ib \in \FSb} \thetah(u_i u_\ib^{\pm})
  }
  \,.
\end{equation}
For the right-hand side, we can trade the sum over the residues at the
poles where
$\{z'_\ib\}_{\ib=1}^{n_2} \subset
\{w_\gamma^{\pm1}\}_{\gamma=1}^{N_3}$ for $(-1)^{n_2}$ times the sum
over the residues at the poles where
$\{z'_\ib\}_{\ib=1}^{n_2} \subset
\{u_\alpha^{\pm1}\}_{\alpha=1}^{N_1}$.  The contribution from the
poles where
$\{z'_\ib\}_{\ib=1}^{n_2} \subset \{u_\ib^{\pm1}\}_{\ib \in \FSb}$ is
precisely given by the sum \eqref{eq:(0,1)-SOR}, and the desired
equality \eqref{eq:(0,1)-triality} follows.

\appendix

\section{The \texorpdfstring{$(0,2)$}{(0,2)} triality}
\label{sec:triality}

Let $N_1$, $N_2$, $N_3$ be nonnegative integers such that
\begin{equation}
  N = \frac12 (N_1 + N_2 + N_3)
\end{equation}
is also an integer.  Let
\begin{equation}
  n_\bullet = N - N_\bullet \,,
  \quad
  \bullet = 1, 2, 3 \,.
\end{equation}
The $(0,2)$ triality \cite{Gadde:2013lxa} is the infrared equivalence
among a triplet of gauge theories $T_{N_1, N_2, N_3}$,
$T_{N_2, N_3, N_1}$ and $T_{N_3, N_1, N_2}$.

The theory $T_{N_1, N_2, N_3}$ is a two-dimensional $\CN = (0,2)$
supersymmetric gauge theory with gauge group $\U(n_3)$ and has
fundamental chiral multiplets $\Phi_\alpha$, $\alpha = 1$, \dots,
$N_1$, antifundamental chiral multiplets $P_\beta$, $\beta = 1$,
\dots, $N_2$, antifundamental Fermi multiplets $\Psi_\gamma$,
$\gamma = 1$, \dots, $N_3$, and $N_1 N_2$ singlet Fermi multiplets
$\Gamma_{\alpha\beta}$.  They are coupled by a vector multiplet for
$\U(n_3)$ and the superpotential
$\sum_{\alpha,\beta} \Phi_\alpha \Gamma_{\alpha\beta} P_\beta$.  The
rank of the gauge group is fixed by the condition for the absence of
nonabelian gauge anomaly.  Furthermore, a pair of determinant Fermi
multiplets $\Omega_+$, $\Omega_-$ is introduced to cancel abelian
gauge anomaly.

The flavor symmetry of $T_{N_1, N_2, N_3}$ is
$\U(N_1) \times \U(N_2) \times \U(N_3) \times \U(1)_\Omega$, under
which the matter multiplets transform as summarized in
Table~\ref{tab:triality}.  We use $K_1^\alpha$, $K_2^\beta$,
$K_3^\gamma$, $K_\Omega$ respectively to denote the Cartan generators
of $\U(N_1)$, $\U(N_2)$, $\U(N_3)$, $\U(1)_\Omega$ such that
\begin{equation}
  K_1^\alpha(\Phi_{\alpha'}) = -\delta^\alpha_{\alpha'} \,,
  \quad
  K_2^\beta(P_{\beta'}) = \delta^\beta_{\beta'} \,,
  \quad
  K_3^\gamma(\Psi_{\gamma'}) = \delta^\gamma_{\gamma'} \,,
  \quad
  K_\Omega(\Omega_\pm) = 1 \,.
\end{equation}
The generator $\Jb_0$ of the $\U(1)$ R-symmetry in the infrared
$\CN = (0,2)$ superconformal algebra can be identified by the
$c$-extremization procedure and is given by
\begin{equation}
  \Jb_0(\Phi_\alpha) = \frac{n_2}{N} \,,
  \quad
  \Jb_0(P_\beta) = \frac{n_1}{N} \,,
  \quad
  \Jb_0(\Gamma_{\alpha\beta}) = \frac{n_3}{N} \,,
  \quad
  \Jb_0(\Psi_\gamma) = \Jb_0(\Omega_\pm) = 0 \,.
\end{equation}

For our purposes it is convenient to shift the R-charge $\Jb_0$ with
Cartan generators of the flavor symmetry so that we can choose the
values of the resulting R-charge $R$ arbitrarily for $\Phi_\alpha$,
$P_\beta$, $\Psi_\gamma$ and $\Omega_+$:
\begin{equation}
  \label{eq:R}
  R
  =
  \Jb_0
  - \sum_\alpha \Bigl(r_\alpha - \frac{n_2}{N}\Bigr) K_1^\alpha
  + \sum_\beta \Bigl(s_\beta - \frac{n_1}{N}\Bigr) K_2^\beta
  + \sum_\gamma t_\gamma K_3^\gamma
  + \Bigl(q_+ - \sum_\gamma t_\gamma\Bigr) K_\Omega \,.
\end{equation}
The R-symmetry $\U(1)_R$ generated by $R$ generically commutes only
with the maximal torus of the flavor symmetry.  The absence of mixed
$\U(1)_R$--gauge anomaly requires
\begin{equation}
  \sum_\alpha (r_\alpha - 1) - \sum_\beta (s_\beta - 1) + \sum_\gamma t_\gamma
  - q_+ + q_-
  = 0 \,,
\end{equation}
which fixes the R-charge $q_-$ of $\Omega_-$.

\begin{table}
  \begin{tabular}{c|cccccccc}
    \hline\hline
    & $\U(n_3)$ & $\U(N_1)$ & $\U(N_2)$ & $\U(N_3)$ & $\U(1)_\Omega$ & $\U(1)_R$
    \\
    \hline
    $\Phi_\alpha$ & $\mathbf{n_3}$ & $\overline{\mathbf{N_1}}$ & $\mathbf{1}$ & $\mathbf{1}$ & $0$ & $r_\alpha$
    \\
    $P_\beta$ & $\overline{\mathbf{n_3}}$ & $\mathbf{1}$ & $\mathbf{N_2}$ & $\mathbf{1}$ & $0$ & $s_\beta$
    \\
    $\Psi_\gamma$ & $\overline{\mathbf{n_3}}$ & $\mathbf{1}$ & $\mathbf{1}$ & $\mathbf{N_3}$ & $0$ & $t_\gamma$
    \\
    $\Gamma_{\alpha\beta}$ & $\mathbf{1}$ & $\mathbf{N_1}$ & $\overline{\mathbf{N_2}}$ & $\mathbf{1}$ & $0$ & $1 - r_\alpha - s_\beta$
    \\
    \hline
    $\Omega_+$ & $\det$ & $\mathbf{1}$ & $\mathbf{1}$ & $\det$ & $1$ & $q_+$
    \\
    $\Omega_-$ & $\det^{-1}$ & $\det$ & $\det$ & $\mathbf{1}$ & $1$ & $q_-$
    \\
    \hline\hline
  \end{tabular}

  \caption{The theory $T_{N_1,N_2,N_3}$.}
  \label{tab:triality}
\end{table}

There are various pieces of evidence for the $(0,2)$ triality.  The
most convincing among them is the agreement of the elliptic genera
across the three theories.  Here we demonstrate that the elliptic
genera in the NS--NS sectors of $T_{N_1,N_2,N_3}$ and
$T_{N_3,N_1,N_2}$ match.

The elliptic genus in the NS--NS sector of an $\CN = (0,2)$
superconformal field theory $T$ with $\U(1)$ flavor symmetries
$\U(1)_\alpha$, $\alpha = 1$, $\dotsc$, $N_f$, is defined by
\begin{equation}
  I_T^{\text{NSNS}}(u_1, \dotsc, u_{N_f}; q)
  =
  \Tr_{\CH_T^{\text{NSNS}}}\biggl(
  (-1)^F q^{L_0} \qb^{\Lb_0 - \Jb_0/2} \prod_\alpha u_\alpha^{K_\alpha}
  \biggr) \,,
\end{equation}
where $L_0$, $\Lb_0$ are generators of the Virasoro algebra,
$K_\alpha$ is a generator of $\U(1)_\alpha$ and $u_\alpha$ is a
fugacity for $K_\alpha$.  The trace is taken in the Hilbert space of
states $\CH_T^{\text{NSNS}}$ in the radial quantization of the NS--NS
sector of the theory.  Due to supersymmetry only states with
$\Lb_0 = \Jb_0/2$ contribute, hence the elliptic genus is independent
of $\qb$.  Using the angular momentum $L_0 - \Lb_0$ the above
expression can be written as
\begin{equation}
  I_T^{\text{NSNS}}(u_1, \dotsc, u_{N_f}; q)
  =
  \Tr_{\CH_T^{\text{NSNS}}}\biggl(
  (-1)^F q^{L_0 - \Lb_0 + \Jb_0/2} \prod_\alpha u_\alpha^{K_\alpha}
  \biggr)
  \,.
\end{equation}
By appropriate rescaling of the flavor fugacities the R-charge $\Jb_0$
entering this expression can be modified to any linear combination of
the form $\Jb_0 + \sum_\alpha c_\alpha K_\alpha$.

When a theory has a standard Lagrangian description, we can write down
its NS--NS elliptic genus using the theta function \cites{Gadde:2013wq,
  Gadde:2013lxa}
\begin{equation}
  \theta(z) = (z;q) (q/z;q) \,,
  \qquad
  (z;q) = \prod_{k=0}^\infty (1 - zq^k) \,.
\end{equation}
We have
\begin{equation}
  \theta(qz) = \theta(1/z) = -\frac{1}{z} \theta(z)
\end{equation}
and
\begin{equation}
  \Res_{z = w} \frac{(q;q)^2}{z \theta((z/w)^{\pm1})}
  =
  \mp1 \,.
\end{equation}

Let $u_\alpha$, $v_\beta$, $w_\gamma$, $x$ be the fugacities for
$K_1^\alpha$, $K_2^\beta$, $K_3^\gamma$, $K_\Omega$, respectively.  The elliptic
genus of $T_{N_1,N_2,N_3}$ is given by the integral
\begin{equation}
  \label{eq:EG-T123}
  I_{T_{N_1,N_2,N_3}}(u,v,w,x)
  =
  \frac{1}{n_3!} \oint \CI_{T_{N_1,N_2,N_3}}(z;u,v,w,x)
  \,,
\end{equation}
where
\begin{multline}
    \CI_{T_{N_1,N_2,N_3}}(z; u,v,w,x)
    \\ \qquad
    =
    \prod_{i} -\frac{\rmd z_i}{2\pi\iu z_i}
    (q;q)^2
    \prod_{i \neq j}
    \theta\biggl(\frac{z_i}{z_j}\biggr)
    \prod_{i,\alpha}
    \theta\biggl(q^{\frac12 r_\alpha} \frac{z_i}{u_\alpha}\biggr)^{-1}
    \prod_{i,\beta}
    \theta\biggl(q^{\frac12 s_\beta} \frac{v_\beta}{z_i}\biggr)^{-1}
    \hfill
    \\ \qquad\quad
    \times
    \prod_{i,\gamma}
    \theta\biggl(q^{\frac12(1+t_\gamma)} \frac{w_\gamma}{z_i}\biggr)
    \prod_{\alpha,\beta}
    \theta\biggl(q^{\frac12(2 - r_\alpha - s_\beta)} \frac{u_\alpha}{v_\beta}\biggr)
    \hfill
    \\ \qquad\quad
    \times
    \theta\biggl(
    q^{\frac12(1 + q_+)} x \prod_{i} z_i \prod_\gamma w_\gamma
    \biggr)
    \theta\biggl(
    q^{\frac12(1 + q_-)} x
    \frac{\prod_\alpha u_\alpha \prod_\beta v_\beta}{\prod_{i} z_i}
    \biggr)
    \,.
    \hfill
\end{multline}
For each variable $z_i$, $i = 1$, \dots, $n_3$, the integrand is a
meromorphic differential on the elliptic curve $\C^\times/q\Z$.  The
domain of integration is chosen in such a way that the integral
receives contributions only from the iterated residues at the poles
coming from the factor
$\prod_{i,\alpha} \theta(q^{\frac12 r_\alpha} z_i/u_\alpha)^{-1}$.

The poles contributing to the integral \eqref{eq:EG-T123} are in
one-to-one correspondence with the ordered choices of $n_3$ elements
from the set $\{1, \dotsc, N_1\}$.  For a given choice, there are
$n_3!$ permutations of elements and the corresponding poles contribute
equally.  Thus, the elliptic genus is a sum over the subsets of
$\{1, \dotsc, N_1\}$ with $n_3$ elements:
\begin{equation}
  I_{T_{N_1,N_2,N_3}}(u,v,w,x)
  =
  \sum_{\substack{\FS \subset \{1, \dotsc, N_1\} \\ |\FS| = n_3}}
  I_{T_{N_1,N_2,N_3}}^\FS(u,v,w,x)
  \,.
\end{equation}
The summands are given by
\begin{multline}
    I_{T_{N_1,N_2,N_3}}^\FS(u,v,w,x)
    \\
    =
    \prod_{i \in \FS} \prod_{\ib \in \FSb}
    \theta\biggl(q^{\frac12(r_{\ib} - r_i)} \frac{u_i}{u_{\ib}}\biggr)^{-1}
    \prod_{i \in \FS} \prod_\gamma
    \theta\biggl(q^{\frac12(1+t_\gamma + r_i)} \frac{w_\gamma}{u_i}\biggr)
    \prod_{\alpha,\beta}
    \theta\biggl(q^{\frac12(r_{\ib} + s_\beta)} \frac{v_\beta}{u_{\ib}}\biggr)
    \\
    \times
    \theta\biggl(
    q^{\frac12(1 + q_+ - \sum_{i \in \FS} r_i)} x
    \prod_{i \in \FS} u_i \prod_\gamma w_\gamma
    \biggr)
    \theta\biggl(
    q^{\frac12(1 + q_- + \sum_{i \in \FS} r_i)} x
    \prod_{\ib \in \FSb} u_{\ib} \prod_\beta v_\beta
    \biggr)
    \,,
  \end{multline}
where the index $i$ now runs over the elements of $\FS$, whereas $\ib$
runs over the $n_2$ elements of the complement $\FSb$ of
$\FS \subset \{1, \dotsc, N_1\}$.

Trading the sum over $\FS$ for a sum over $\FSb$, we can rewrite the
elliptic genus as a similar integral with $n_2$ integration variables,
\begin{equation}
  I_{T_{N_1,N_2,N_3}}(u,v,w,x)
  =
  \frac{(-1)^{n_2}}{n_2!} \oint
  \CI_{T_{N_3,N_1,N_2}}(z'; u,v,w,x)
  \,,
\end{equation}
where the integrand
\begin{multline}
  \CI_{T_{N_3,N_1,N_2}}(z'; u,v,w,x)
    \\
    =
    \prod_{\ib} -\frac{\rmd z'_{\ib}}{2\pi\iu z'_{\ib}}
    (q;q)^2
    \prod_{\ib \neq \jb}
    \theta\biggl(\frac{z'_{\ib}}{z'_{\jb}}\biggr)
    \prod_{{\ib},\gamma}
    \theta\biggl(q^{\frac12(1-t_\gamma - \Delta)} \frac{z'_{\ib}}{w_\gamma}\biggr)^{-1}
    \prod_{{\ib},\alpha}
    \theta\biggl(q^{\frac12(-r_\alpha + \Delta)} \frac{u_\alpha}{z'_{\ib}}\biggr)^{-1}
    \hfill
    \\ \quad
    \times
    \prod_{{\ib},\beta}
    \theta\biggl(q^{\frac12(s_\beta + \Delta)} \frac{v_\beta}{z'_{\ib}}\biggr)
    \prod_{c,\alpha}
    \theta\biggl(q^{\frac12(1 + t_\gamma + r_\alpha)} \frac{w_\gamma}{u_\alpha}\biggr)
    \hfill
    \\ \quad
    \times
    \theta\biggl(
    q^{\frac12(1 + q_- + \sum_\alpha r_\alpha - n_2 \Delta)} x \prod_{{\ib}} z'_{\ib} \prod_\beta v_\beta
    \biggr)
    \theta\biggl(
    q^{\frac12(1 + q_+ - \sum_\alpha r_\alpha + n_2 \Delta)} x
    \frac{\prod_\alpha u_\alpha \prod_\gamma w_\gamma}{\prod_{{\ib}} z'_{\ib}}
    \biggr) \,,
\end{multline}
is a meromorphic differential on $\C^\times/q^\Z$ for each variable
$z'_{\ib}$, $\ib = 1$, $\dots$, $n_2$, and the domain of integration
encloses the poles from
$\prod_{\ib,\gamma} \theta(q^{\frac12(-r_\alpha + \Delta)}
u_\alpha/z'_{\ib})^{-1}$.  Since the sum of the residues of a
meromorphic differential on a compact Riemann surface vanishes, we can
change the domain to one that picks up the poles from
$\prod_{\ib,\gamma} \theta(q^{\frac12(1-t_\gamma - \Delta)}
z'_{\ib}/w_\gamma)^{-1}$ after dropping the sign factor $(-1)^{n_2}$
in front of the integral.  Doing so precisely yields the integral
formula for the elliptic genus of the theory $T_{N_3,N_1,N_2}$, with
the matter contents summarized in Table~\ref{tab:triality-2}.

The constant $\Delta$ in the R-charge assignment accounts for the
freedom of shifting the R-charge by a multiple of a generator of the
overall $\U(1)$ factor of the gauge group $\U(n_2)$.  Such a shift
acts trivially on physical states and operators.  At the level of the
integral formula for the elliptic genus, the shift corresponds to
rescaling the integration variables by a power of $q$.  If we set
\begin{equation}
  R(\Omega_+) = R(\Omega'_-) \,,
  \qquad
  R(\Omega_-) = R(\Omega'_+) \,,
\end{equation}
then we have
\begin{equation}
  \Delta = \frac{1}{n_2} \sum_\alpha r_\alpha \,.
\end{equation}

\begin{table}
  \begin{tabular}{c|cccccccc}
    \hline\hline
    & $\U(n_2)$ & $\U(N_1)$ & $\U(N_2)$ & $\U(N_3)$ & $\U(1)_\Omega$ & $\U(1)_R$
    \\
    \hline
    $P'_\alpha$ & $\overline{\mathbf{n_2}}$ & $\mathbf{N_1}$ & $\mathbf{1}$ & $\mathbf{1}$ & $0$ & $-r_\alpha + \Delta$
    \\
    $\Psi'_\beta$ & $\overline{\mathbf{n_2}}$ & $\mathbf{1}$ & $\mathbf{N_2}$ & $\mathbf{1}$ & $0$ & $s_\beta - 1 + \Delta$
    \\
    $\Phi'_\gamma$ & $\mathbf{n_2}$ & $\mathbf{1}$ & $\mathbf{1}$ & $\overline{\mathbf{N_3}}$ & $0$ & $1 - t_\gamma - \Delta$
    \\
    $\Gamma'_{\gamma\alpha}$ & $\mathbf{1}$ & $\overline{\mathbf{N_1}}$ & $\mathbf{1}$ & $\mathbf{N_3}$ & $0$ & $t_\gamma + r_\alpha$
    \\
    \hline
    $\Omega'_+$ & $\det$ & $\mathbf{1}$ & $\det$ & $\mathbf{1}$ & $1$ & $q_- + \sum_\alpha r_\alpha - n_2 \Delta$
    \\
    $\Omega'_-$ & $\det^{-1}$ & $\det$ & $\mathbf{1}$ & $\det$ & $1$ & $q_+ - \sum_\alpha r_\alpha + n_2 \Delta$
    \\
    \hline\hline
  \end{tabular}

  \caption{The theory $T_{N_3,N_1,N_2}$ related to $T_{N_1,N_2,N_3}$
    by the $(0,2)$ triality.}
  \label{tab:triality-2}
\end{table}

\section*{Acknowledgment}

This work is supported by NSFC Grant Number 12375064.

\end{document}